\documentclass[aip,jcp,reprint,floatfix]{revtex4-1} 
\usepackage[utf8]{inputenc}

\usepackage{natbib}
\usepackage{graphicx}
\usepackage{xcolor}
\usepackage{amsmath}
\usepackage{float}
\usepackage[margin=2.5cm]{geometry} 
\usepackage{physics}
\usepackage{amssymb}
\usepackage{pifont}

\newcommand{\eps}{\epsilon}
\renewcommand{\Tr}{\mathrm{Tr}}

\newcommand{\nocc}{n_\mathrm{occ}}
\newcommand{\nvir}{n_\mathrm{vir}}
\newcommand{\nphys}{n_\mathrm{phys}}
\newcommand{\naux}{n_\mathrm{aux}}

\newcommand{\nmom}{n_\mathrm{mom}}
\newcommand{\nqmo}{n_\mathrm{QMO}}

\newcommand{\fext}{F_\mathrm{ext}}
\newcommand{\eoneb}{E_\mathrm{1b}}
\newcommand{\etwob}{E_\mathrm{2b}}
\newcommand{\emptwo}{E_\mathrm{MP2}}
\newcommand{\ecorr}{E_\mathrm{corr}}

\newcommand{\toadd}[1]{#1}
\newcommand{\toremove}[1]{}

\begin{document} 

\title{A wave function perspective and efficient truncation of renormalised second-order perturbation theory}
\author{Oliver J. Backhouse}
\affiliation{Department of Physics, King's College London, Strand, London WC2R 2LS, U.K.}
\author{Max Nusspickel}
\affiliation{Department of Physics, King's College London, Strand, London WC2R 2LS, U.K.}
\author{George H. Booth}
\email{george.booth@kcl.ac.uk}
\affiliation{Department of Physics, King's College London, Strand, London WC2R 2LS, U.K.}


\begin{abstract}
We present an approach to renormalized second-order Green's function perturbation theory (GF2) which avoids all dependency on continuous variables, grids or explicit Green's functions, and is instead formulated entirely in terms of static quantities and wave functions. Correlation effects from MP2 diagrams are iteratively incorporated to modify the underlying spectrum of excitations by coupling the physical system to fictitious auxiliary degrees of freedom, allowing for the single-particle orbitals to delocalize into this additional space. The overall approach is shown to be rigorously $\mathcal{O}[N^5]$, after an appropriate compression of this auxiliary space. This is achieved via a novel scheme which ensures that a desired number of moments of the underlying occupied and virtual spectra are conserved in the compression, allowing a rapid and systematically improvable convergence to the limit of the effective dynamical resolution. The approach is found to then allow for the qualitative description of stronger correlation effects, avoiding the divergences of MP2, as well as its orbital-optimized version.
On application to the G1 test set, we find that modifications to only up to the third spectral moment of the underlying spectrum from which the double excitations are built is required for accurate energetics, even in strongly correlated regimes. This is beyond simple self-consistent changes to the density matrix of the system, but far from requiring a description of the full dynamics of the frequency-dependent self-energy.
\end{abstract}
\maketitle

\section{Introduction}
Mean-field methods such as Hartree--Fock (HF) and density functional theory (DFT) are cost-effective solutions to describe a wealth of quantum chemical and condensed matter properties, however their single-determinant formalism generally fails in their description of stronger correlation effects, as well as other important physical phenomena such as dispersion interactions, reactive intermediates or charge transfer states\cite{doi:10.1021/cr200107z}. This difference is manifest in the failure of DFT to describe electron correlation accurately through its approximate exchange-correlation functional. For stronger correlation, multireference models such as the complete active space self-consistent field (CASSCF)\cite{ROOS1980157} and multireference perturbation theories\cite{doi:10.1063/1.1515317, doi:10.1002/qua.23052, doi:10.1063/1.4928643} can be used to quantify static correlation, however the number of determinants in the wave function grows exponentially with the size of the active space, and therefore renders them rapidly intractable\cite{KNOWLES1984315}. Whilst emerging approaches such as the density matrix renormalization group\cite{PhysRevB.48.10345, doi:10.1146/annurev-physchem-032210-103338, doi:10.1063/1.4976644} and quantum Monte Carlo\cite{doi:10.1063/1.3193710, doi:10.1021/acs.jctc.5b00917, doi:10.1063/1.5055390} techniques are bringing down this cost, it is still an expensive option, largely restricted to small molecules.
On the other hand, methods based on single-determinant reference generally achieve polynomial scaling and provide a range of approximations to the correlation energy. The dichotomy between cost and accuracy underpins much of the modern research into the simulation of quantum chemistry and condensed matter\cite{doi:10.1021/bk-2007-0958.ch005}.

One of the main tools in this regard is perturbation theory, where the traditional starting point is to partition the Hamiltonian into a mean-field part $\hat{H}_{0}$ and remainder describing the explicit two-particle interaction potential, $\hat{V}$. Rayleigh--Schr{\"o}dinger perturbation theory on this partitioning results in the M{\o}ller-Plesset (MP$n$) class of methods, most commonly performed to second-order (MP2)\citep{Moller1934}. The MP2 method is the simplest way of including electron correlation post-HF and scaling with system size is a $\mathcal{O}[N^5]$ prior to many effective cost-reduction schemes, allowing its application to large and condensed phase systems, whilst correctly describing dispersion and pair correlation effects\cite{doi:10.1063/1.4940732, doi:10.1063/1.3072903, doi:10.1063/1.3466765, booth13}. Despite its success, MP2 does not account for any inter-pair electron correlation even qualitatively, which manifest in a number of situations such as strongly correlated problems, molecules away from equilibrium or screening. This is well illustrated in molecular dissociation events, where the restricted MP2 energy rapidly diverges with increasing separation as strong correlation takes hold\cite{doi:10.1021/bk-2007-0958.ch005}. Furthermore, the restriction of MP2 to uncoupled pair correlation renders its ultimate accuracy limited, and can be hard to justify in many contexts compared to the computationally cheaper DFT approach. Higher orders MP$n$ will capture much of this neglected correlation and can delay the divergence when stretching bonds, however the increased scaling with system size as well as potential non-convergence of the series overall\cite{doi:10.1063/1.472352,doi:10.1063/1.481611} gives favor to resummed methods such as coupled cluster singles and doubles (CCSD)\citep{Cizek1966,RevModPhys.79.291}, which generally performs better at a scaling of $\mathcal{O}[N^6]$. 

This resumming of low-order diagrams often allows for many-electron (more than two-body) physics to be qualitatively described via coupled two-body processes, as can be seen by expansion of the exponential form of the wave function ansatz in CCSD. Whilst still non-variational, it does not suffer from the divergence of MP2 in molecular dissociation events for single bonds, however will still fail in instances of substantial strong correlation. Including a subset of third-order diagrams as a perturbation constitutes CCSD(T)\cite{RAGHAVACHARI1989479}, which is widely considered to be the gold standard in quantum chemistry, however scales as $\mathcal{O}[N^7]$. A challenge remains to efficiently and self-consistently couple the bare pair correlation physics of MP2 together, whilst still retaining its favorable $\mathcal{O}[N^5]$ scaling, to give rise to a set of renormalized diagrams which contains at least some portion of many-body, inter-pair correlations present in \toadd{e.g.} CCSD. Achieving this could admit higher accuracy results and allow a wider scope of applicability of the method in the presence of stronger correlation effects. An approach within this framework is described in this work.


In MP2, the $\hat{H}_{0}$ can be considered to be fully defined by the orbitals and energies resulting from the diagonalization of the Fock matrix, describing the kinetic energy as well as the mean-field Coulomb and exchange contributions of each electron. This is manifest in the mean-field uncorrelated spectrum of the system. Via Koopman's theorem, each orbital energy can be considered an approximation to its ionization potential (occupied orbitals) or electron affinity (unoccupied orbitals). However, this spectrum neglects orbital relaxation, as well as all contributions from correlated physics, which would modify (at times substantially) the true ionization potential and electron affinities, and hence the underlying spectrum. One possible self-consistent renormalization of MP2 diagrams therefore involves updating the spectrum, to take into account the correlated effects in its description, and hence redefining this zeroth-order description of the system. This modifies the electron and hole propagators to take into account the effects of MP2 correlation into account.

One approach to modify this underlying spectrum and $\hat{H}_0$ is the orbital-optimized MP2 method (OO-MP2), which iteratively optimizes the orbitals (and hence the electron density and spectrum) to relax them in the presence of the MP2 correlations\citep{Bozkaya2011,Bozkaya2014}. This is done by minimizing the Hylleraas functional and ensuring the energy is stationary with respect to the orbital rotations in a variational fashion in response to the MP2 correlations. However, this does not allow a full relaxation of the spectrum, or even density matrix, since the zeroth-order system is still described in a simple orbital picture. For instance, the relaxed density matrix is still constrained to be idempotent, which is not a property of a density matrix resulting from a true correlated system. Nevertheless, the OO-MP2 energies can improve upon MP2 results in situations of moderate correlation while still at $\mathcal{O}[N^5]$ scaling, as well as the stationarity with respect to orbital rotations simplifying subsequent property computation. However, the divergences of MP2 are also observed for OO-MP2, motivating parameterized MP2 methods to combat such deficiencies. These include employing level-shifting\cite{Stuck2013}, and $\kappa$-OO-MP2\cite{Joonho2018,Joonho2019}, the latter similar to the single-reference driven similarity renormalization group (DSRG)\cite{Evangelista2014}. Further techniques to empirically improve upon MP2 energetics without increasing cost include the spin-component scaled (SCS-MP2)\cite{Grimme2003} and spin-opposite scaled (SOS-MP2)\cite{Yousung2004} variants, which has also been combined with orbital-optimization\cite{Lochan2007,Neese2009,Bozkaya2014b}.

The computational difficulties in describing a true correlated density matrix is similar to the challenges in defining a fully correlated spectrum for the optimized $\hat{H}_{0}$ , which is impossible within the constraints of simply adjusting the mean-field potential or keeping a simple single-particle orbital framework, and so a different formulation is required. This is most commonly done within a Green's function formalism, whereby the spectrum is defined according to a continuous function of energy (or its conjugate variable, time). This allows for arbitrary changes to this underlying spectrum due to the correlation at the level of MP2 via a self-energy, until self-consistency is achieved, defining the `second-order Green's function perturbation theory' (GF2)\cite{Holleboom1990,VanNeck2001,Dahlen2005,Phillips2014}. This approach has been investigated in the literature, and shown that it can prevent divergences in MP2, as the correlated physics can renormalize the spectrum, opening the bandgap to allow for convergent results even in strong correlation limits\cite{doi:10.1063/1.4940900,PhysRevB.100.085112}. Furthermore, the `flat-plane' condition requiring piece-wise linearity in the energy of a system as their electron number or spin-polarization is changed is well captured within the approach, demonstrating a substantial reduction in strong correlation and self-interaction error compared to most approximate DFT functionals\cite{doi:10.1063/1.4921259}. However, GF2 is substantially more complicated than the MP2 method conceptually, computationally and algorithmically. This stems primarily from having to efficiently discretize the continuous (time and/or frequency) variable, as well as to perform Fourier transforms between them in order to achieve an efficient algorithm and optimal $\mathcal{O}[N^5]$ scaling. This choice of discretization is critical in reducing the large prefactor to these calculations, and substantial work has been done in choosing these grids optimally\cite{doi:10.1021/acs.jctc.6b00178, PhysRevB.98.075127}.

In this work, we will consider an alternate framework to systematically truncate the GF2 approach to resum low-order MP2 diagrams to infinite order, whilst retaining an $\mathcal{O}[N^5]$ scaling. This approach allows for its recasting back in terms of wave functions, and admits an efficient, zero-temperature sum-over-states formalism, more akin to traditional MP2 formulations. It avoids the construction of any explicitly frequency or time-dependent quantities, discretization of continuous grids, or numerical Fourier transforms. Instead, the modifications necessary to self-consistently update the effective propagators of $\hat{H}_{0}$ in the required manner involve the construction of an auxiliary space in which this zeroth-order Hamiltonian additionally spans. \toadd{The concept of the auxiliary space has previously been used in diagrammatic perturbation theory, for example in the algebraic diagrammatric construction (ADC) method\cite{Rusakov2016}.} This truncation of the full GF2 approach is then rationalized as a self-consistency on the occupied and virtual \textit{moments} of the spectrum, ensuring that these are appropriately renormalized due to the correlations of MP2. It will be seen in numerical results that this systematic truncation rapidly approaches the full GF2 results, applied to both the dissociation of molecules where strong correlation effects dominate\cite{doi:10.1021/bk-2007-0958.ch005}, as well as the weaker correlation effects of the G1 molecular test set\cite{doi:10.1063/1.3008061}, and improving the description of intermediately correlated systems compared to the native MP2 method. The method also bears similarities to other implementations of GF2 which are performed purely in the frequency-domain\cite{VanNeck2001}, but which use other approximation techniques for compression, and we will compare to these as well as more traditional iterative `orbital-optimized' MP2 approaches (OO-MP2)\cite{Bozkaya2011}.

\section{Background: Second-order Green's function perturbation theory} \label{sec:GF2}
We start with a short recap of the salient features of the GF2 method, as this informs and motivates the approximations developed in subsequent sections. However, this is not essential to understand the subsequent method. For further details on efficient and complete implementations of the GF2 method which this work builds on, we refer the reader to Refs.~\cite{Dahlen2005,Phillips2014}. 

GF2 is a renormalized pair perturbation theory which is cast exclusively in the language of Green's functions and self-energies. This allows for renormalization of the propagator lines of diagrams due to the effects of correlation, which cannot be achieved in the simple molecular orbital picture.
The self-energy is a dynamical, single-particle object which describes the effects required to change the Green's function and spectrum of the system, which in this case is due to the presence of correlation in the form of the MP2 direct and exchange second-order diagrams. The effect of this self-energy allows for changes in the spectrum of the physical system, such as adding additional states and/or shifting and suppressing the weight of states from the original HF description. In this way, the mean-field result can be made to better represent the correlated nature of the system, and the propagation of the real electrons or holes through it. In GF2, this is used to update the propagators, which in turn redefine the effect of the MP2 second-order diagrams, and this is iteratively performed to self-consistency, summing subsets of diagrams to infinite order via renormalization of the underlying propagators of the system.

The Green's function and (irreducible) self-energy are linked via the Dyson equation
\begin{equation} \label{eq:dyson}
    G(\omega) = G_{0}(\omega) + G_{0}(\omega) \Sigma(\omega) G(\omega),
\end{equation}
which represents an infinite series providing all possible insertions of the diagrams in $\Sigma$ into the propagator. $G_{0}$ represents the zeroth-order (mean-field) Green's function, defined as
\begin{equation}\label{eq:zero_greens_function}
    G_{0}(\omega) = [\omega I - F]^{-1},
\end{equation}
where $F$ is the generalized physical Fock matrix, with elements given by
\begin{equation} \label{eq:Fock}
    F_{pq} = h_{pq} + \sum_{rs} [(pq|rs) - \frac{1}{2}(ps|rq)] D_{rs},
\end{equation}
where $h$ is the `core' kinetic-energy and nuclear-electron Hamiltonian, $D_{rs}$ is the one-body density matrix of the system, $(pq|rs)$ denotes standard `chemist notation' two-electron repulsion integrals, and the summation is over spatial orbitals of the full system. In GF2, Eq.~\ref{eq:dyson} is used to iteratively renormalize the Green's function of the system with the self-energy. Subsequent contributions to the self-energy are generally computed in the time-domain after Fourier transform of the resulting Green's function, as
\begin{align} \label{eq:TimeSE}
    \Sigma_{pq}(\tau)=-\sum_{rstuvw}  G_{rs}(\tau)G_{tu}(\tau)G_{vw}(-\tau)  \nonumber \\
                     (pt|wr)[2(qu|vs)-(qs|vu)] ,
\end{align}
where corresponding changes to the frequency-independent part of the self-energy (first-order diagrams) are found via the changes to the density matrix and update of Eq.~\ref{eq:Fock}.
At convergence, the self-energy contains the second-order direct and exchange diagrams encountered in MP2, renormalized \textit{ad infinitum}, resulting in a conserving, rigorously diagrammatic approximation, and correlated spectrum and one-body density matrix of the system\cite{Dahlen2005}.

The Green's function and self-energy of GF2 rely on frequency and time grids which must be carefully controlled to minimize numerical errors, while the scaling in the formulation above ensures that the prefactor to the most expensive step (evaluation of Eq.~\ref{eq:TimeSE}) is dependent on the resolution of the time grid, $N_\tau$ ($\mathcal{O}[N^{5}N_\tau]$ overall). The requirement of a Fourier transform between these two physical quantities also means that the use of non-uniform quadrature is non-trivial, but intermediate representations to develop efficient algorithms are a source of recent and ongoing research\cite{doi:10.1021/acs.jctc.6b00178,PhysRevB.98.075127}. Furthermore, when one uses the Matsubara (imaginary frequency) axis to ensure a smooth and efficient frequency-domain Green's function representation, it is necessarily a finite temperature representation. This can be an advantage for thermal properties, but results in increasing cost to converge to zero-temperature in systems with a small spectral gap. \toadd{As a result, obtaining the spectral function requires analytic continuation of the Matsubara Green's function onto the real axis, which is an ill-conditioned problem, plagued by numerical artifacts and substantial loss of precision at higher energies.}

We now turn to a framework which removes the need for a definition of a grid in either domain, by instead defining a zeroth-order {\em static} Hamiltonian, which spans an additional fictitious subspace required to rigorously define the effects of an implicit self-energy at zero temperature. This auxiliary space is in turn defined via a projection in order to match the moments of the spectral distribution of the occupied and unoccupied states up to a given order. Explicit self-energies and Green's functions can be constructed if desired in the approach, which will converge to traditional GF2 quantities, and connections will be made between the formalism presented and analogous dynamical quantities extensively. \toadd{This framework also has the advantage of allowing one to obtain the spectral function without the need for analytic continuation. The accuracy of this direct real-frequency spectrum, will be explored more in a future study.}



\section{Auxiliary-GF2}\label{sec:agf2}
In the following approach, we mimic the effects of correlation on the renormalized spectrum and propagators via coupling the system to an external, fictitious set of degrees of freedom, rather than an explicit self-energy. The delocalization of the resulting single-particle orbitals into this additional space can implicitly model the same effects as a fully dynamical self-energy. This additional space is comprised of `auxiliary' states, and while they can have one-electron contributions to the single-particle Hamiltonian (both amongst themselves, and connecting to the original `physical' states) they do not have any two-body interactions spanning this space. Appropriate choice of these Hamiltonian matrix elements renormalizes the propagators of the physical space due to the effect of the second-order correlations. To distinguish this approach from the traditional, grid-based `dynamical' formulation detailed in Sec.~\ref{sec:GF2}, we will denote the approach {\em Auxiliary-GF2} (AGF2). We present the spin-free working equations for the restricted algorithm, where spin symmetry breaking is not allowed. However, the formalism is simply adapted to work within an unrestricted spin-orbital basis where spin-breaking is possible, and we denote the restricted and unrestricted cases as RAGF2 and UAGF2, respectively.

\begin{figure}
    \centering
    \includegraphics[scale=0.5]{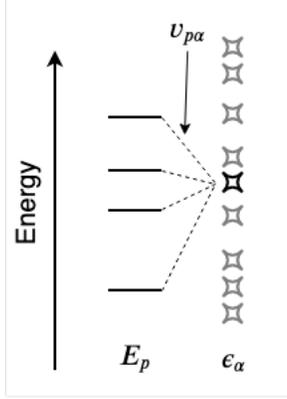}
    \caption{Diagram showing the coupling of a set of physical orbitals (in this case Hartree--Fock orbitals) coupled to a set of auxiliaries with energy $\epsilon_{\alpha}$ and a coupling strength $v_{p \alpha}$. Only the coupling for one such auxiliary is shown, with that auxiliary shown in bold.}
    \label{fig:aux_phys_coupling}
\end{figure}

To make this analogy between a set of fictitious states and a self-energy more rigorous, we can write a manifestly causal self-energy in the frequency domain as
\begin{equation}\label{eq:aux_self_energy}
    \Sigma_{pq}(\omega) = \sum_{\alpha}^{\naux} \frac{v_{p \alpha} v^{\dagger}_{q \alpha}}{\omega - \epsilon_{\alpha}},
\end{equation}
where $\omega$ is a continuous frequency variable, which can equally well be defined on the real or imaginary (Matsubara) axis. Defining $p, q$ to be the original Hartree--Fock orbitals spanning the physical orbital space of the system, the $\naux$ auxiliary states $\alpha$, can be defined as coupled via the one-electron Hamiltonian matrix elements, $v_{p \alpha}$, and having one-electron energies given by $\epsilon_{\alpha}$. Each auxiliary state therefore contributes a single pole in the form of the overall self-energy. This auxiliary space coupling is expressed schematically in Fig.~\ref{fig:aux_phys_coupling}, with the notation of the indices used defined in Tab.~\ref{tab:indices}.

\begin{table}
    \centering
    \begin{tabular}{l l}
        \hline 
        Orbital space & Indices \\
        \hline
        General HF (physical) orbitals & $p, q, r, s \dots$ \\
        Auxiliary orbitals & $\alpha, \beta, \gamma, \delta \dots$ \\
        Occupied `quasi'-molecular orbitals \quad & $i, j, k, l \dots$ \\
        Virtual `quasi'-molecular orbitals & $a, b, c, d \dots$ \\
        General `quasi'-molecular orbitals & $w, x, y, z \dots$ \\
        \hline
    \end{tabular}
    \caption{Notation for indexing of different orbital spaces used in this work.}
    \label{tab:indices}
\end{table}

Given a set of auxiliary Hamiltonian elements $v_{p \alpha}$ and $\epsilon_{\alpha}$, an {\em extended} Fock matrix can be constructed, $\fext$, which defines a single-particle eigenvalue problem over the original physical Hartree--Fock and additional auxiliary space, as
\begin{equation}\label{eq:fock_ext}
    \begin{bmatrix}
    & F_\mathrm{phys} & v & \\
    & v^\dagger & \mathrm{diag}(\epsilon_\mathrm{aux}) &
    \end{bmatrix}
    \phi = \lambda \phi,
\end{equation}
where $F_\mathrm{phys}$ is the physical space Fock matrix and $\epsilon_\mathrm{aux}$ is the vector of $\epsilon_{\alpha}$ auxiliary energies. The diagonalization of this sparse $\fext$ matrix results in the definition of `quasi'-molecular orbitals (QMOs), denoted $\phi$, which are single-particle states which diagonalize $\fext$ and span the union of both the physical and auxiliary space. These QMOs now define a propagator in the physical space which has been renormalized by the effects of a self-energy defined as Eq.~\ref{eq:aux_self_energy}, and therefore is equivalent to the effect of Dyson's equation as applied directly to the zeroth-order Green's function as given in Eq.~\ref{eq:dyson}. This can be seen by
defining the Green's function as
\begin{multline}
    \begin{bmatrix}
    & \omega - F_\mathrm{phys} & v & \\
    & v^\dagger & \omega - \epsilon_{\mathrm{aux}} &
    \end{bmatrix}
    \begin{bmatrix}
    & G_\mathrm{phys} & G_\mathrm{phys,aux} & \\
    & G_\mathrm{aux,phys} & G_\mathrm{aux} &
    \end{bmatrix} \\ 
    = 
    \begin{bmatrix}
    & I & 0 & \\
    & 0 & I &
    \end{bmatrix}. \\
\end{multline}
Solving the equations above for the projection of the resulting Green's function in the physical space, $G_\mathrm{phys}$, it can be shown that this extended Fock matrix construction is equivalent to the introduction of a fully dynamical self-energy via Dyson's equation in Eq.~\ref{eq:dyson} as
\begin{align}
    G_\mathrm{phys}(\omega) &= \left( \omega I - F_\mathrm{phys} - v (\omega I - \epsilon_\mathrm{aux})^{-1} v^\dagger \right)^{-1} \label{eq:1stderiv} \\
    &= (\omega I - F_\mathrm{phys} - \Sigma(\omega) )^{-1} \\
    &= (G_0(\omega)^{-1} - \Sigma(\omega) )^{-1} \\
    &= G_0(\omega) + G_0(\omega) \Sigma(\omega) G_\mathrm{phys}(\omega) \label{eq:lastderiv}.
\end{align}

Once these QMOs have been constructed via diagonalization of the extended Fock matrix of Eq.~\ref{eq:fock_ext}, the filling of the QMOs is determined such that the number of electrons in the physical space matches the desired number of electrons. The number of electrons in the physical space, $N_{\mathrm{phys}}$ is given by the trace of the projection of the density matrix of the occupied QMOs into the physical space ($D_{pq}$), as
\begin{align}
    N_\mathrm{phys} &= \Tr[{D_{pq}}] \\
    &=\Tr\left[2{\sum_i^{n_{\mathrm{occ}}^{\mathrm{QMO}}} \phi_{pi} \phi^*_{qi}}\right] \label{eq:Dphys} \\
    &= 2 \sum_p^{n_{\mathrm{phys}}} \sum_i^{n_{\mathrm{occ}}^{\mathrm{QMO}}} \phi_{pi} \phi^*_{pi}  .
\end{align}
If $N_\mathrm{phys}$ is not identically equal to the desired number of electrons in the physical space, then a small chemical potential, $\mu$ is added to the auxiliary space to shift their energies relative to the physical space, and therefore finely control the number of electrons in the physical space. This can be done by iteratively diagonalizing the resulting extended Fock matrix, and optimizing $\mu$ to give the desired $N_{\textrm{phys}}$ number of electrons, e.g. by bisection. Furthermore, the resulting change in the description of the physical space can also result in renormalization of the first-order diagrams contained in the Coulomb and exchange terms of the Fock matrix. In order to include these effects, the updated physical projection of the density matrix is also used in order to recompute the exchange and Coulomb terms in the $F_\mathrm{phys}$ block of the extended Fock matrix, and this also iterated to convergence. To accelerate this convergence, as in self-consistent field calculations, we employ the direct inversion of the iterative subspace (DIIS)\cite{doi:10.1002/jcc.540030413}.

\subsection{Definition of auxiliary space parameters} \label{sec:aux_params}

We now consider how the parameters defining the auxiliary space are to be defined (i.e. the matrix of one-body physical-auxiliary couplings $v$, and the vector of auxiliary energies, $\epsilon_\mathrm{aux}$, of Eq.~\ref{eq:fock_ext}), in order to rigorously include the second-order diagrammatic effects we are after. We first consider the definition of the frequency-domain MP2 self-energy, $\Sigma_{\mathrm{MP2}}(\omega)$, which can be defined as the Fourier transform of Eq.~\ref{eq:TimeSE} in the case of a diagonal $G(\tau)$, as
\begin{align}\label{eq:mp2_self_energy}
    [\Sigma_{\mathrm{MP2}}(\omega)]_{pq}
    &= 
    \sum_{ij}^{\nocc} \sum_{a}^{\nvir} \frac{(pi|ja)[2(qi|ja)-(qj|ia)]}{\omega+E_{a}-E_{i}-E_{j}}
    \\
    &+
    \sum_{i}^{\nocc} \sum_{ab}^{\nvir} \frac{(pa|bi)[2(qa|bi)-(qb|ai)]}{\omega+E_{i}-E_{a}-E_{b}}.
    \nonumber
\end{align}
Here $p,q$ represent all physical degrees of freedom in the original Hartree--Fock basis, ensuring that the self-energy spans only the physical space. In the first iteration, the $i,j$ and $a,b$ indices represent canonical occupied and virtual orbitals, with energies $E$, with the algorithm giving the traditional MP2 self-energy. 

The aim is now to represent this self-energy as a set of frequency-independent auxiliary orbitals, by recasting Eq.~\ref{eq:mp2_self_energy} in the form of Eq.~\ref{eq:aux_self_energy}. This is shown in Eqs.~\ref{eq:1stderiv} - \ref{eq:lastderiv} to have an equivalent effect on the propagator of the physical space, and therefore allow the desired iterative renormalization of the effective particle dynamics. To do this, the numerator of the rearranged form of Eq.~\ref{eq:mp2_self_energy} must be able to be written as a physical space matrix valued operator formed from outer-products of vectors. Taking only the occupied part (first term) of Eq.~\ref{eq:mp2_self_energy}, one may exploit the symmetry in the summations to rearrange its form as
\begin{widetext}
\begin{align}
    \label{eq:pole_deriv}
    [&\Sigma_\mathrm{MP2}^\mathrm{occ}(\omega)]_{pq}
    = 
    \sum_{ij}^{\nocc} \sum_{a}^{\nvir} \frac{(pi|ja)[2(qi|ja)-(qj|ia)]}{\omega+E_{a}-E_{i}-E_{j}}
    \\
    &= 
    \sum_{i<j}^{n_\mathrm{occ}} \sum_{a}^{n_\mathrm{vir}} \frac{(pi|ja) [2 (qi|ja) - (qj|ia)] + (pj|ia) [2 (qj|ia) - (qi|ja)]}{\omega + E_a - E_i - E_j} 
    +
    \sum_{i}^{n_\mathrm{occ}} \sum_{a}^{n_\mathrm{vir}} \frac{(pi|ia) [2 (qi|ia) - (qi|ia)]}{\omega + E_{a}-2E_{i}}
    \\
    &= \sum_{i<j}^{n_\mathrm{occ}} \sum_{a}^{n_\mathrm{vir}} \frac{3[(pi|ja) - (pj|ia)] [(qi|ja) - (qj|ia)] + [(pi|ja) + (pj|ia)] [(qi|ja) + (qj|ia)]}{2(\omega + E_{a}-E_{i}-E_{j})} 
    +
    \sum_{i}^{n_\mathrm{occ}} \sum_{a}^{n_\mathrm{vir}} \frac{(pi|ia)(qi|ia)}{\omega + E_{a}-2E_{i}}.
    \label{eq:pole_deriv_last}
\end{align}
\end{widetext}
Exploiting the permutational symmetry of the two-electron integrals, the form of Eq.~\ref{eq:pole_deriv_last} allows for the casting of each term onto an auxiliary orbital in the extended Fock matrix to mimic its effect, as dictated by Eq.~\ref{eq:aux_self_energy}. All states have the correct outer product form, ensuring that all effects of the exchange and Coulomb diagrams are rigorously causal, which is not guaranteed for other diagrammatic theories\cite{PhysRevB.90.115134}. Each unique set of two occupied and one virtual MO indices maps to two degenerate auxiliary orbitals. An equivalent rearrangement applies to the virtual (particle) part of Eq.~\ref{eq:mp2_self_energy}, with the auxiliary energies and couplings given in Table~\ref{tab:poles}. An equivalent derivation exists when one starts with a spin-unrestricted self-energy. These auxiliaries number $\nocc \nvir \nphys$ and therefore the number of auxiliaries scales as $\mathcal{O}[\nphys ^{3}]$.

\begin{table}
\centering
\begin{tabular}{c|c|c}
    \hline
    $\epsilon_\alpha$ & $v_{p\alpha}$ & Constraint \\
    \hline
    $E_i + E_j - E_a$ & $\frac{\sqrt{6}}{2} [(pi|ja) - (pj|ia)]$ & $i < j$ \\
    $E_i + E_j - E_a$ & $\frac{\sqrt{2}}{2} [(pi|ja) + (pj|ia)]$ & $i < j$ \\
    $2E_i - E_a$ & $(pi|ia)$ & \\
    $E_a + E_b - E_i$ & $\frac{\sqrt{6}}{2} [(pa|bi) - (pb|ai)]$ & $a < b$ \\
    $E_a + E_b - E_i$ & $\frac{\sqrt{2}}{2} [(pa|bi) + (pb|ai)]$ & $a < b$ \\
    $2E_a - E_i$ & $(pa|ai)$ & \\
    \hline
\end{tabular}
\caption{Summary of auxiliary energies ($\eps_{\alpha}$) and couplings to physical orbitals ($v_{p \alpha}$) required to build extended Fock matrix.}
\label{tab:poles}
\end{table}

In order to iterate the effects of this correlation and include its effect on the propagator, after these auxiliaries have been included, the orbitals denoted by the indices $i,j,a,b$ above are replaced by the eigenstates of the extended Fock matrix. These therefore represent the occupied and virtual quasi-molecular orbitals (QMOs), spanning both the physical and auxiliary spaces defined in Sec.~\ref{sec:agf2}. This means that the Hartree--Fock energies of Eq.~\ref{eq:pole_deriv_last} and Tab.~\ref{tab:poles} ($E$) are replaced with the eigenvalues of Eq.~\ref{eq:fock_ext}, $\lambda$, which are the energies of the QMOs. Furthermore, the auxiliary coupling strengths also need to be modified. This requires a three-quarter transformation of the two-electron repulsion integrals (ERIs) into the QMO basis to be performed each iteration. Note that the leading index remains in the physical Hartree--Fock basis, and therefore does not need to be transformed. We transform these integrals once each iteration, via three quarter-transforms, as
\begin{equation}\label{eq:integral_transform}
    (pi|ja) = \sum_{qrs}^{\nphys} \phi_{qi} \phi_{rj} \phi_{sa} (pq|rs) ,
\end{equation}
where $(pq|rs)$ are the original ERIs in the Hartree--Fock basis. Note that (as with MP2 theory) this integral transform constitutes the leading-order computationally scaling step, scaling as \mbox{$\mathcal{O}[\nphys^2(\nphys + \naux)^3]$}, although the prefactor can be minimized by realising that only certain occupied-virtual combinations are required. The efficiency of the algorithm is therefore significantly impacted by the size of the auxiliary space, which must scale as $\nphys$ or less to ensure that the overall scaling is not greater than MP2.

However, it can be seen from Table~\ref{tab:poles} that each iteration, the number of auxiliary states grows as $\mathcal{O}[N_{\mathrm{QMO}}^3]$.  Unchecked, this would result in a growth in the size of the auxiliary space as a function of iteration, $t$, as $\mathcal{O}[\nphys^{3^t}]$, which is clearly unphysically high. As a result, a key step in the algorithm is to robustly and effectively compress the information in the auxiliary space, which is described in the next section.

\subsection{Compression of the auxiliary space} \label{sec:compression_schemes}
The scaling of the auxiliary space in the algorithm above is artificial. Even at convergence of all properties, further iterations will continue to grow the size of the auxiliary space, which must necessarily be entirely redundant. The number of auxiliary states will also rapidly exceed the total number of auxiliaries required to represent the exact self-energy, without approximation, which can only have at most the dimensionality of the $(N+1)$- and $(N-1)$-electron FCI space. Furthermore, we know that grid-based Green's function methods are successful precisely because the dynamical information in a self-energy can be effectively discretized in values on a grid in some domain, generally with an {\em a priori} imposed energetic structure from this distribution of grid points which compresses their information in a non-exponentially scaling fashion\cite{doi:10.1021/acs.jctc.6b00178}. In this section, we consider ways to reduce the number and scaling of the auxiliary states created each iteration, from $\mathcal{O}[N_{\mathrm{QMO}}^3]$ down to at most $\mathcal{O}[\nphys]$. 

This resulting scaling can be rigorously rationalized on physical grounds, and will also be shown numerically. The self-energy has an energy-dependence and spatial-dependence. In the formalism above, each auxiliary describes the spatial entanglement at a specific energy, given by the $\epsilon_{\alpha}$ energy of the auxiliary orbital. In large systems, the energy scales required to be spanned should not increase with the size of the system. If this were true, then the size and resolution of the grids used to describe traditional self-energies would also have a system size dependence -- instead, the overall required energy resolution of the propagator rapidly saturates with system size, and therefore scales as $\mathcal{O}[1]$. Furthermore, constraints such as standard finite basis sets ensure that we are always probing an effective low-energy model which does not significantly increase the energy scale as the system gets larger. This leaves the spatial dependence, which can be characterized by a matrix of couplings $v$, which increases {\em at most} linearly with the size of the physical space. If the number of auxiliaries at a given energy were larger than the number of physical degrees of freedom, then a singular value decomposition of their couplings would isolate the non-null space of auxiliaries coupling to the system at that energy, and would necessarily be no more than the number of physical states.

This overall linear scaling in the number of auxiliaries to represent {\em any} self-energy does not therefore rely on effects such as spatial locality, and should hold regardless of the physical character of the physical states denoted by $p,q$ in Sec.~\ref{sec:aux_params}. However, it would likely be possible in large system limits to asymptotically approach $\mathcal{O}[1]$ scaling of the size of the auxiliary space. This is because in a local basis, the spatial structure of the self-energy would become diagonally-dominated as physically distant degrees of freedom would have vanishing weight of second-order self-energy diagrams connecting them, allowing for a natural reduction in the scaling of the auxiliaries required to describe this structure. Similar locality approximations have also reduced scaling for MP2 methods\cite{doi:10.1063/1.3072903,doi:10.1063/1.4940732}. All equations and derivations in Sec.~\ref{sec:aux_params} naturally transfer to a local basis, however in this work we represent the physical space in canonical Hartree--Fock orbitals, and therefore do not exploit efficiency savings due to spatial locality, which we save for future work.

The redundancy in the auxiliary space arises due to auxiliary states which are both close in energy, and which have a high degree of overlap in their couplings to the physical space characterized in $v$. There exist approaches to compactify the auxiliary space with heuristic algorithms based on this criterion, e.g. Ref.~\onlinecite{PhysRevB.90.085102}. In other methods such as dynamical mean-field theory (DMFT) it is common to numerically fit a compact set of auxiliaries (known as bath states in DMFT). The parameters of these bath states result from the numerical minimization of some measure of error in this effective self-energy compared to the full self-energy defined on a grid (generally the hybridization function, defined on the imaginary-frequency axis). While this is reasonable for small physical spaces, this approach returns to the requirement of (and dependence on) a numerical grid representation of the self-energy, while the numerical fit on the Matsubara axis necessitates a loss of accuracy, and becomes increasingly difficult and rapidly intractable for larger systems with many bath states\cite{Liebsch_2011,mejutozaera2019efficient}.

We instead consider two approaches below to compress the auxiliary space, which are outlined, followed by a discussion of their efficiency, accuracy and the overall auxiliary compression algorithm used in this work.

\subsubsection{Consistency in self-energy moments}\label{sec:semom}
In the work of Ref.~\onlinecite{VanNeck2001}, Van Neck et al. develop a GF2 algorithm with a similar grid-free, discrete pole self-consistency, applied to a series of small atomic systems\citep{VanNeck2001,Piers2002}. In their work however, the Dyson equation was solved directly and numerically, rather than by diagonalization of the extended Fock matrix of auxiliaries as is performed here. Nevertheless, there was still a requirement of a scheme to reduce the number of explicit poles (c.f. auxiliaries) of the resulting self-energy.
In order to do this, they adopted the BAsis GEnerated by Lanczos (BAGEL) method, first developed for this purpose in the nuclear physics community\citep{Muther1988,Muther1993,Dewulf1997}, and based on a block Lanczos tridiagonalization of the extended Fock matrix\citep{Kim1989,Iguchi1992}. This method provides a compression scheme which is capable of reducing a (potentially large) set of discrete self-energy poles, such that the resulting compressed poles spans a basis which ensures a matching of a number of moments of the original occupied (hole) and unoccupied (particle) self-energy. We denote the truncation of the (separate hole and particle) self-energy as $\nmom^{\Sigma}$, which ensures that these self-energy moments are matched to an order of $2\nmom^{\Sigma}+1$.
As an example, $\nmom^{\Sigma}=0$ ensures that the integrated weight over all frequencies and mean for the separate hole and particle self-energy (for each element) is maintained before and after compression (i.e. $v v^{\dagger}$ and $v \epsilon v^{\dagger}$). Increasing $\nmom^{\Sigma}$ will match higher moments, covering the variance, skew and higher order moments of the self-energy distribution of each element. 

This compression whilst conserving these moments is achieved using the block Lanczos tridiagonalisation algorithm. The method proceeds by applying the block Lanczos algorithm to $\fext$, and truncating the number of Lanczos iterations to $\nmom^{\Sigma}+1$. This converts the large $\fext$ matrix into a smaller block-tridiagonal matrix, $F_\mathrm{block}$ of the form
\begin{equation}\label{eq:fock_ext_tridiag}
    F_\mathrm{block} = 
    \begin{bmatrix}
    F_\mathrm{phys} & T_{1} &        & 0 \\
    T_{1}^{\dagger} & M_{1} & T_{2}  & \\
          & T_{2}^{\dagger} & M_{2}  & \ddots \\
    0     &       & \ddots & \ddots \\
    \end{bmatrix},
\end{equation}
which consists of $(\nmom^{\Sigma}+1)$ on- and off-diagonal matrices ($M_{i}$ and $T_{i}$ respectively), each of dimension $\nphys$. One may then diagonalise the external subspace of this matrix, where the eigenvalues constitute the new, compressed auxiliary energies ($\epsilon_{\alpha}$), while the matrix of couplings to the new auxiliary space ($v_{p \alpha}$) can be obtained as $\sum_{\beta} (T_{1})_{p \beta} C_{\beta \alpha}$, where $C$ is the matrix of eigenvectors spanning the external subspace. This process is applied separately to the occupied (hole) and virtual (particle) parts of the auxiliary space. The number of auxiliaries which result from this process is $2 \nphys (\nmom^{\Sigma}+1)$, independent of the initial number of discrete auxiliaries. The computational complexity of this compression of an auxiliary space is $\mathcal{O}[n_\mathrm{aux} \nphys^2 \nmom^{\Sigma} + (\nmom^{\Sigma} \nphys)^3]$, where the first term corresponds to the cost of the block Lanczos iterations, while the second corresponds to the resulting diagonalization of the $F_\mathrm{block}$ external space, and $n_\mathrm{aux}$ here denotes the number of auxiliaries before compression\footnote{Additionally, when working in finite-precision arithmetic, one may require a reorthogonalization step in order to maintain orthogonality of the Lanczos vectors. This introduces a further $\mathcal{O}[\nphys^{2} \naux]$ cost step, but does not affect the overall scaling.}.

\subsubsection{Consistency in Green's Function moments}\label{sec:gfmom}
Inspired by the energy-weighted density matrix embedding theory (EwDMET)\citep{Fertitta2018,Fertitta2019} there also exists a formulation to (non-iteratively) construct a compact auxiliary space, which instead conserves the separate hole and particle moments of the resulting {\em Green's function}, rather than the self-energy. Therefore, the full action of the entire original auxiliary space (the full dynamics of the effective self-energy) is considered, and it is the effect on the resulting Green's function which is truncated. The moments are defined in a similar way, and we will denote the maximum truncation by a parameter, $\nmom^G$. Similar to the self-energy moment truncation, a truncation to order $\nmom^G$ exactly conserves the moments of the resulting particle and hole physical-space Green's function to an order of $2\nmom^G+1$ due to a Wigner $2n+1$ rule of perturbation theory\footnote{$\nmom^G$ refers to the order of the response of the ground state to the auxiliary space, which is why the order of moments is conserved to $2\nmom^G+1$.}. This means that $\nmom^G=0$ maintains the separate particle and hole integrated weight and mean for each $G_\mathrm{phys}$ element before and after compression of the auxiliary space (i.e. $\phi \phi^{\dagger}$ and $\phi \lambda \phi^{\dagger}$ in the physical space, in
contrast to Sec.~\ref{sec:semom}), while higher moments conserve increasing resolution of this distribution in energy. These moments also represent physical expectation values (unlike the self-energy) which can be computed as
\begin{align}
    H_{pq}^{(n)}=\langle \Phi | c_q^{\dagger}[c_p,\fext]_{\{n\}}| \Phi \rangle \\
    P_{pq}^{(n)}=\langle \Phi | [c_p,\fext]_{\{n\}} c_q^{\dagger} | \Phi \rangle
\end{align}
where $H_{pq}^{(n)}$ and $P_{pq}^{(n)}$ represent the $n^{\mathrm{th}}$ moments of the separate hole and particle (physical) Green's function respectively, $|\Phi\rangle$ represents the determinant of occupied QMOs, and $[c_p,\fext ]_{\{n\}}=[\dots[[c_p,\fext],\fext],\dots \fext]$ represents $n$-nested commutators, with $[c_p,\fext ]_{\{0\}}=c_p$. Note that all $\nmom^G$ thresholds result in a compressed auxiliary set which nevertheless exactly reproduce the physical space 1-body density matrix coupled to the full set of auxiliaries, ensuring (amongst other things) that the number of physical electrons is exactly preserved.

However, in order to construct the reduced auxiliary space which represents these moments faithfully, it is necessary to completely diagonalize $\fext$ spanning the physical and uncompressed auxiliary spaces, to obtain the QMOs as in Eq.~\ref{eq:fock_ext}. The projection to the compressed auxiliary space then proceeds by building a set of contracted auxiliaries which are linear combinations of the original $\naux$ auxiliary space, for each order $0 \leq n \leq \nmom^G$, separately for the occupied and virtual part of the Green's function. This results in a set of $2 \nphys$ contracted auxiliary states per order, given by
\begin{align}
    \label{eq:gfmom_vector_occ}
    | \chi_{p}^{(n), \mathrm{occ}} \rangle &= 
    \sum_{\alpha}^{\naux} \sum_i^{n_{\mathrm{occ}}^{\mathrm{QMO}}} \phi_{p i} \lambda_i^n \phi_{\alpha i}^* | \alpha \rangle \\
    \label{eq:gfmom_vector_vir}
    | \chi_{p}^{(n), \mathrm{virt}} \rangle &=
    \sum_{\alpha}^{\naux} \sum_a^{n_{\mathrm{virt}}^{\mathrm{QMO}}} \phi_{p a} \lambda_a^n \phi_{\alpha a}^* | \alpha \rangle,
\end{align}
where $| \alpha \rangle$ denotes states in the original auxiliary space, $p$ labels physical orbitals, and $i,a$ run over occupied and virtual QMOs ($\phi$) from the diagonalization of $\fext$ respectively.
The vectors of Eqs.~\ref{eq:gfmom_vector_occ} and \ref{eq:gfmom_vector_vir} are not orthonormal by default. They are first normalized (as they can become very large), before proceeding to an orthogonalization of the entire set of vectors.
The $n=0$ occupied and virtual states are linearly dependent, and higher order moments may have additional linear dependencies, and therefore orthogonalization allows for the further reduction of the number of these vectors. The auxiliary space of $\fext$ is then projected into this new space, and diagonalized to obtain the resulting energies ($\epsilon$) and couplings ($v$) of this truncated set of auxiliary orbitals, in the same fashion as the self-energy consistent truncation described in Sec.~\ref{sec:semom}.

This approach of truncating the moments of the resulting Green's function has the drawback that it requires a complete diagonalization of the full (uncompressed) extended Fock matrix to compute the complete QMO space, with a scaling of $\mathcal{O}[(\nphys + \naux)^3]$, higher than the self-energy moment truncation. However, the final number of compressed auxiliaries is smaller, with a maximum of $\nphys (2 \nmom + 1)$, but frequently less due to the linear dependencies. This again results in a final number which is independent of the initial number of auxiliary states, and is also linear with the number of physical degrees of freedom, which was rationalised as the appropriate physical scaling of the required number of auxiliary states. We now turn to the accuracy of each approach, and find that a combination of these two algorithms can result in dramatically improved scheme with the best features of both.




\subsubsection{Comparison and combination of compression algorithms} \label{sec:compression_overview}
\begin{table}[H]
    \centering
    \begin{tabular}{l | c | c}
        Truncation & $\mathcal{O}$[cost] & \# Compressed aux. \\ 
        \hline
        $\Sigma_\mathrm{occ}$ / $\Sigma_\mathrm{virt}$ & $\naux n_\mathrm{p}^2 \nmom^{\Sigma} + (\nmom^{\Sigma} n_\mathrm{p})^3$ & $2 n_\mathrm{p} (\nmom^{\Sigma} + 1)$ \\
        $G_\mathrm{occ}$ / $G_\mathrm{virt}$ & $(n_\mathrm{p} + \naux)^3$ & $\leq n_\mathrm{p} (2 \nmom^G + 1)$ \\
    \end{tabular}
    \caption{Summary of the cost and resulting number of compressed auxiliary states in the two compression schemes described. $n_\mathrm{p}$ denotes the number of physical orbitals in the system, $\naux$ denotes the initial (uncompressed) number of fictitious auxiliary states, and $\nmom^{\Sigma} / \nmom^G$ represents the order of the hole/particle self-energy or Green's function truncation respectively in the compression algorithm. Note that the physical number of moments in each particle and hole sector which are conserved is given by $2\nmom+1$.}
    \label{tab:compression_overview}
\end{table}

\begin{figure} 
    \centering
\includegraphics[trim={0 0 0 0},width=1.0\columnwidth]{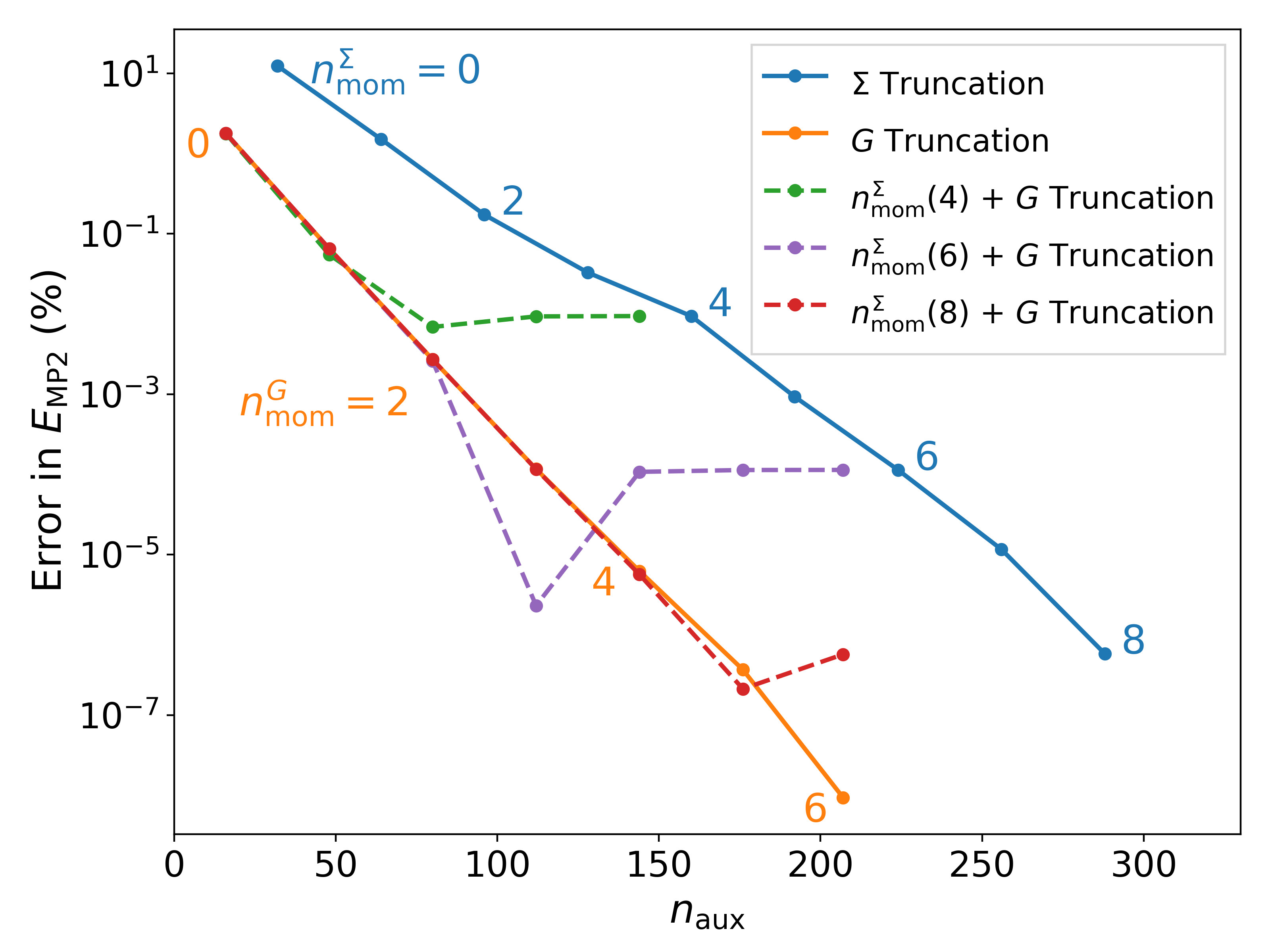}
\includegraphics[trim={0 0 0 0},width=1.0\columnwidth]{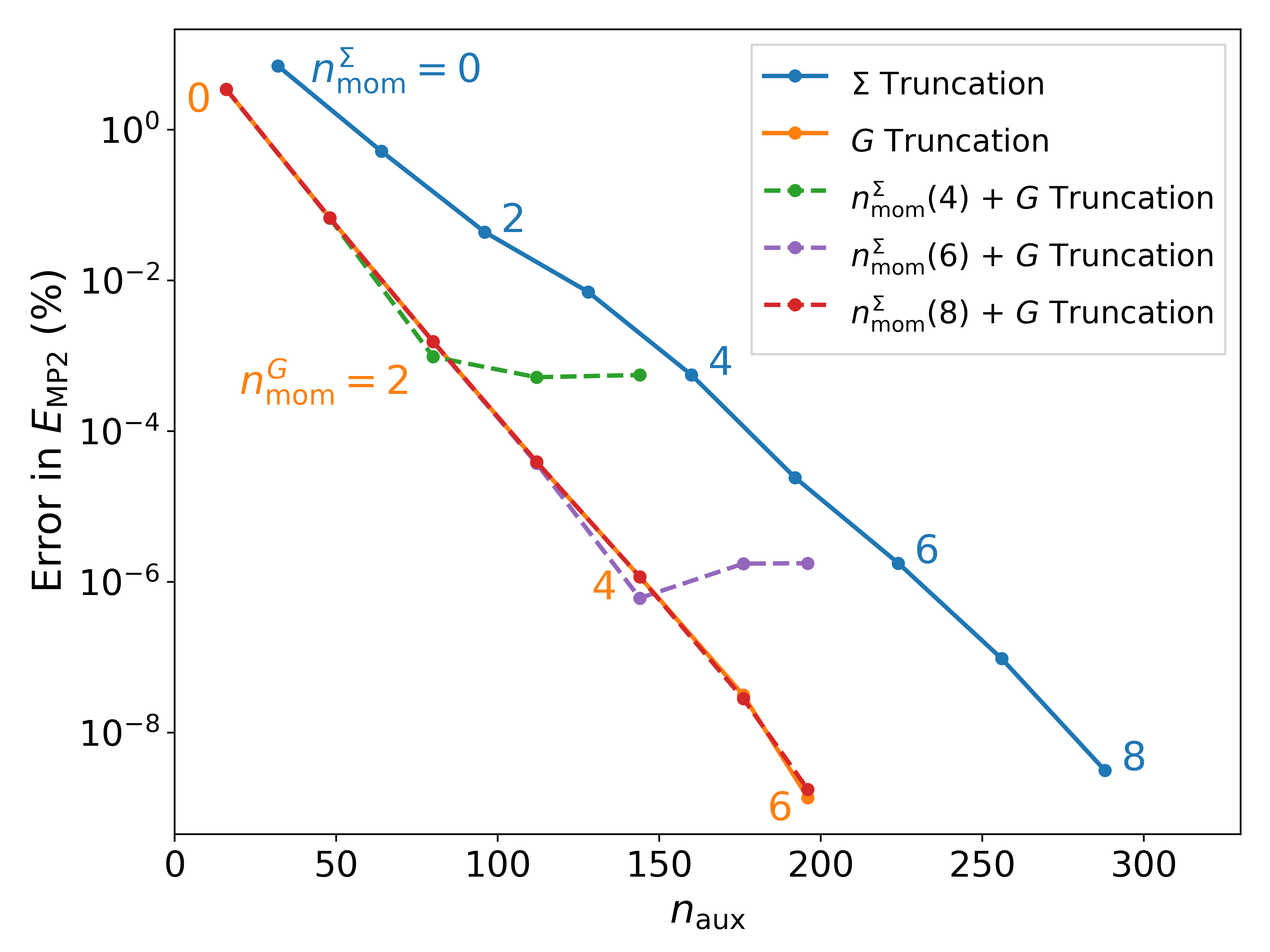}
    \caption{Comparison of the error in the MP2 correlation energy for an H$_{16}$ ring in the STO-3G basis between the two compression schemes based on the matching of moments of the occupied/virtual part of the self-energy \mbox{(`$\Sigma$ Truncation'} in blue) or Green's function \mbox{(`$G$ Truncation'} in yellow). The $x$-axis denotes the number of resulting auxiliary states after the compression. The upper plot shows a weakly correlated system with each bond length being $0.75 \ \mathrm{\AA}$, while the lower plot shows a more strongly correlated system where the bond length is $1.25 \ \mathrm{\AA}$. The initial number of uncompressed auxiliary states in both cases is 796. Numbers on the plot denote the maximum truncation order in each compression scheme, i.e. $\nmom^{\Sigma}$ and $\nmom^G$. Also shown in dotted lines is the error from a hybrid compression scheme, where first the full space of auxiliaries are compressed via truncation of the self-energy moments to a specified order, followed by a further compression to match increasing $\nmom^G$.}
    \label{fig:naux_comparison_h16}
\end{figure}

Table~\ref{tab:compression_overview} summarizes the computational cost and resulting number of compressed auxiliary states, given an initial number $\naux$, for the self-energy and Green's function truncations detailed above. The solid lines in Figure~\ref{fig:naux_comparison_h16} show the resulting error that each of these truncation schemes makes for the MP2 correlation energy in compressing of the auxiliary space in a H$_{16}$ ring system at both an equilibrium and stretched geometry. This represents the first iteration of the self-consistent algorithm above, and we choose to consider just the first (MP2) iteration to avoid mixing effects due to truncation of a single auxiliary space with any effects which may be derived from the self-consistency of this approximation. The initial number of auxiliaries to exactly represent the self-energy in both cases is 796. 

\begin{figure}[H]
    \centering
    \includegraphics[trim={0 0 0 0},width=1.\columnwidth]{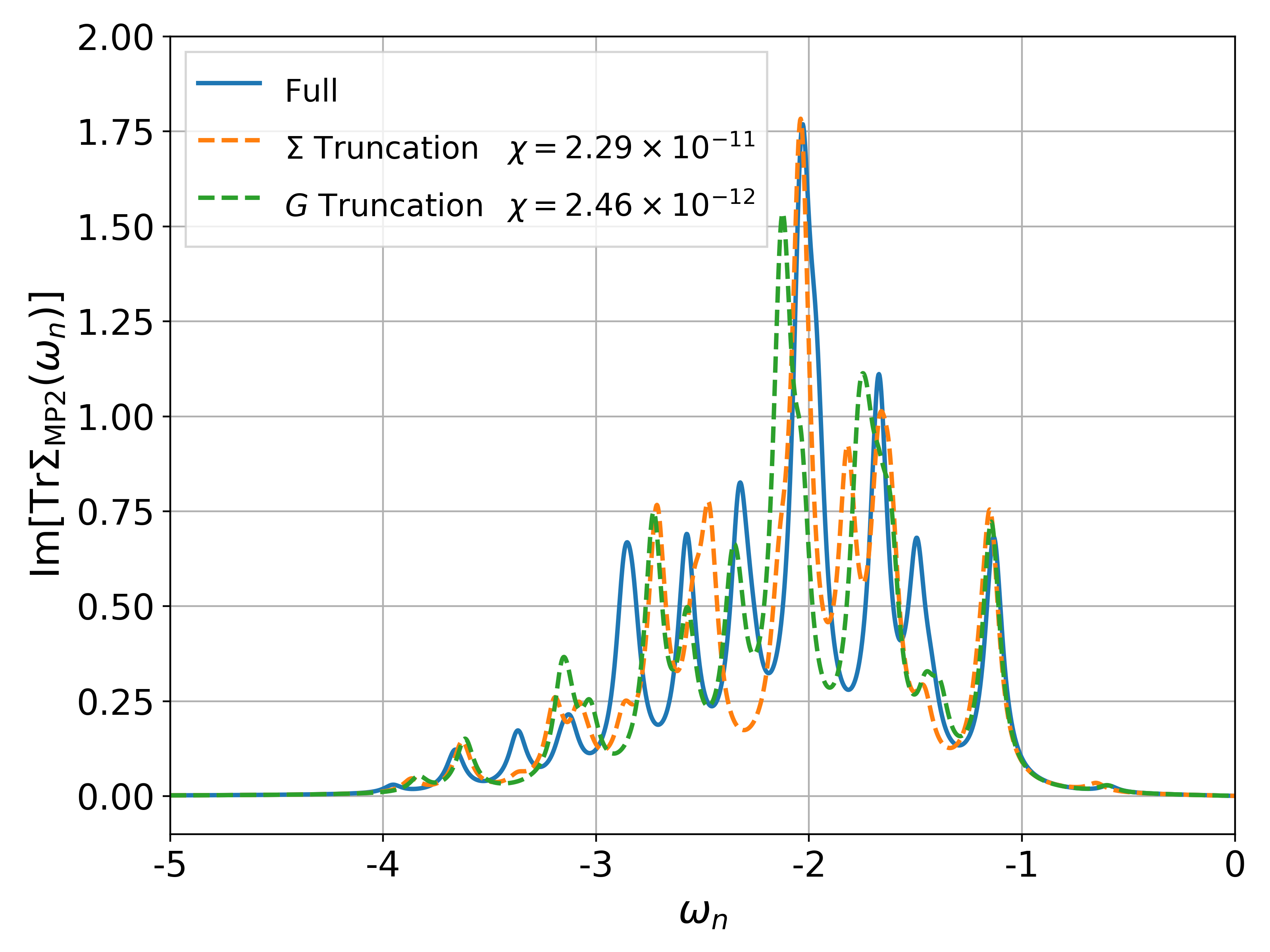}
    \caption{Comparison of the imaginary part of the occupied MP2 self-energy plotted on the real-frequency axis for each compression method, for a H$_{8}$ ring ($r=0.75 \ \mathrm{\AA}$, STO-3G), with a broadening of 50 m$E_\mathrm{h}$. Self-energy and Green's function truncation are both performed to $n_\mathrm{mom}=3$. The error on the Matsubara axis is given as $\chi = \frac{1}{N_{\omega_n}} \sum_{n} \frac{1}{\omega_n} \vert \Sigma(i\omega_n) - \tilde{\Sigma}(i\omega_n) \vert ^{2}$, where $\tilde{\Sigma}(i\omega)$ represents the corresponding compressed auxiliary representation for the self-energy. The value of this error is $2.29\times10^{-11}$ for the $\nmom^{\Sigma}=3$ compression, while only $2.46\times10^{-12}$ for the $\nmom^G=3$ compression.}
    \label{fig:real_se_h8}
\end{figure}

It is found that whilst both compression schemes are systematically improvable, the truncation based on matching of the hole/particle moments of the Green's function gives a substantially more accurate and compact auxiliary representation for the correlation energy, for both correlation strengths. This behaviour is seen to be representative across a number of systems considered, at both weak and stronger correlation strengths. Furthermore, the rate of convergence to the exact result is faster with increasing $\nmom^G$ than $\nmom^{\Sigma}$. We can also consider the effect on the self-energy, rather than total energy, by recasting the effective self-energy on a grid, and comparing to the exact (uncompressed) self-energy. Figure~\ref{fig:real_se_h8} shows the imaginary component of the occupied MP2 self-energy for each compressed scheme (with $\nmom^G = \nmom^{\Sigma} = 3$), now explicitly represented on a real-frequency grid, where contributions to the self-energy from each auxiliary have been broadened by 50 m$E_\mathrm{h}$ as a visual aid. Again, it can be seen that the $\nmom^G$ truncation seems to give a qualitatively improved agreement to the exact result. The agreement is impressive, since these compression schemes are designed to match the moments of the self-energy and Green's function, they do not necessarily preserve the structure of the auxiliaries themselves. Furthermore, this accuracy can be quantified by considering an error function $\chi$ (defined in the figure caption), which is most conveniently evaluated on the imaginary (Matsubara) frequency axis where the self-energy is a smooth function. It can be seen that the Green's function truncation gives almost an order of magnitude reduction in this error estimate compared to the self-energy truncation (from $2.29\times10^{-11}$ to $2.46\times10^{-12}$).

However whilst more accurate, the difficulty in the Green's function auxiliary compression lies in its steeper cost to evaluate, dominated by the cubic scaling with number of original auxiliary states. In the overall algorithm, the two computational bottlenecks are the diagonalization of the extended Fock matrix to form the QMOs ($\mathcal{O}[(\nphys + \naux)^3]$), and more significantly, the three-quarter transformation of the integrals into the QMO basis ($\mathcal{O}[\nphys^2 (\nphys + \naux)^3]$). Furthermore, the number of auxiliary states scales cubically with the number of QMOs in the previous iteration (which is at least the number of physical degrees of freedom, as it is in the first iteration). Therefore, purely using the Green's function auxiliary compression each iteration would result in an overall $\mathcal{O}[\nphys^9]$ scaling algorithm.

Instead, we consider a hybrid two-step auxiliary compression scheme which proves very effective.
In this, the self-energy moment truncation is used to perform an initial reduction to a high $\nmom^{\Sigma}$ moment order, and then the Green's function truncation is applied to this resulting set of auxiliaries to compress them further with a truncation of $\nmom^G$. This ensures that the Green's function truncation cost is independent of the initial number of auxiliaries, and the overall cost for compression is only linear with the number of initial auxiliaries, given by the leading-order scaling of the $\nmom^{\Sigma}$ truncation of $\mathcal{O}[\naux \nphys^2 \nmom^{\Sigma} + (\nphys \nmom^{\Sigma})^3]$. Therefore the compression step scales as $\mathcal{O}[\nphys^5]$ overall. The resulting number of auxiliary states is determined purely by $\nmom^G$ as given in Table~\ref{tab:compression_overview}.


\begin{figure} 
    \centering
    \includegraphics[trim={0 0 0 0},width=1.\columnwidth]{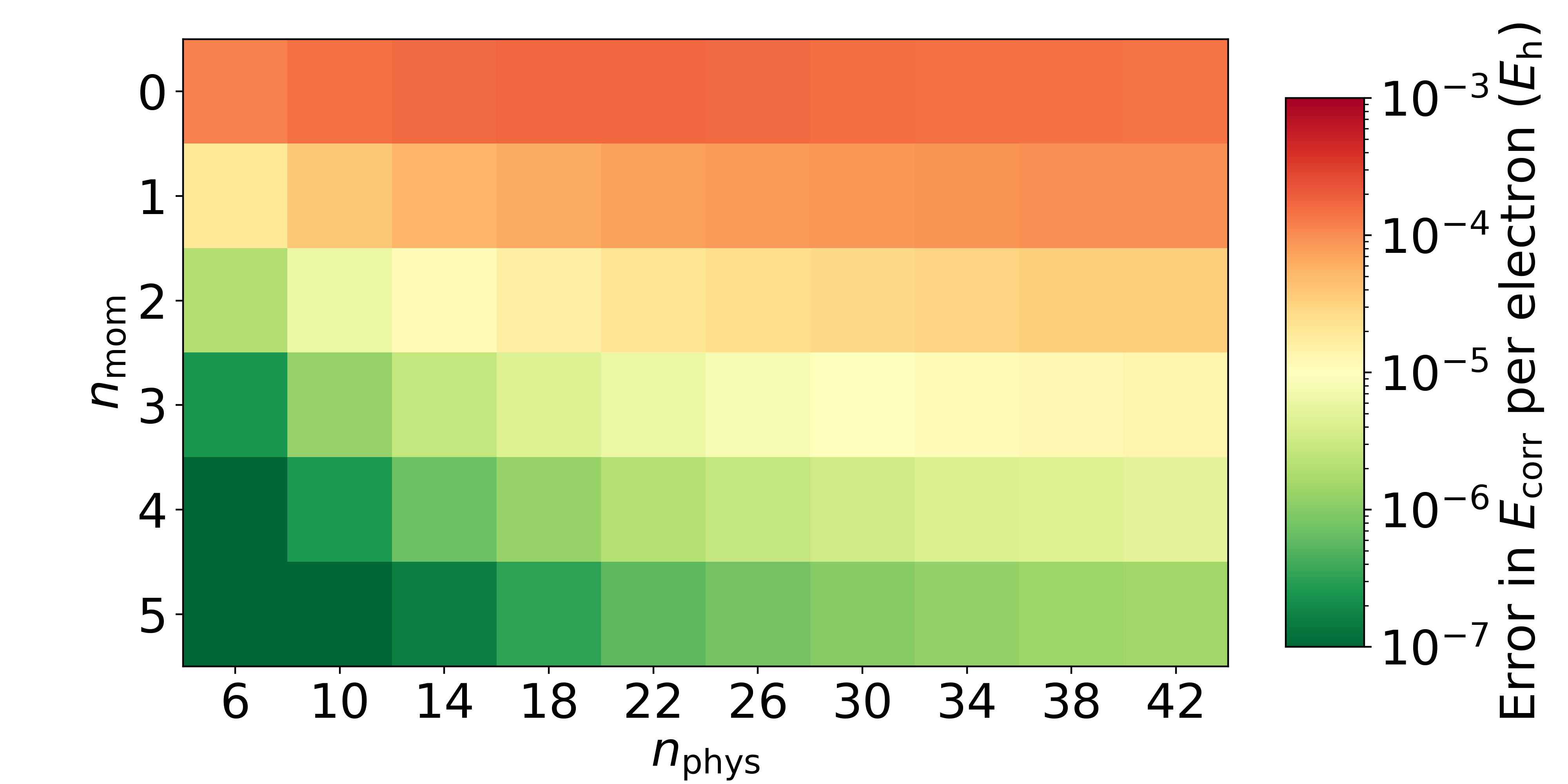}
    \caption{Heat map showing errors in $\ecorr$ per electron for converged AGF2 calculations with different numbers of moments and size of system, for the 
    two-stage AGF2 compression first to $\nmom^{\Sigma}=7$, and then $\nmom^G$. The systems are a series of linear hydrogen chains H$_{n}$ with $R_\mathrm{HH}=1.0 \ \mathrm{\AA}$ in a STO-3G basis, with overall $\nphys$ number of (physical) orbitals. \toadd{For H$_{42}$, there were 42, 126, 210, 294, 378 and 462 auxiliary orbitals for $\nmom^G = 0 \rightarrow 5$, respectively.}}
    \label{fig:hchain_heatmap}
\end{figure}

Figure \ref{fig:naux_comparison_h16} also shows the percentage error in the correlation energy for this lower-scaling, two-step hybrid compression of the auxiliaries. It is found that the initial compression via the self-energy moment matching barely affects the quality of the subsequent Green's function compression, until it reaches the limit of accuracy as defined by the initial $\nmom^{\Sigma}$ truncation at which point increasing $\nmom^G$ does not further improve the accuracy. This behaviour is the same for both weakly and strongly correlated limits, and allows for a significant further reduction in the number of resulting auxiliaries, reducing the leading order step in the algorithm which depends cubically on their number. To demonstrate the scaling in the moment order in a fully self-consistent auxiliary-GF2 calculation, in Fig.~\ref{fig:hchain_heatmap} we demonstrate the absolute errors at convergence for a 
$\nmom^{\Sigma}=7$ followed by $\nmom^G$ compression. This is done across a range of linear hydrogen chains of increasing size H$_{n}$, where $n=4i+2$ for $i \in \mathbb{Z}$, demonstrating the scaling of the moment expansion as the system increases in size. Errors are per-electron and taken with respect to a converged large-moment calculation, which was found easier and more robust to converge to the fully dynamical limit than a traditional grid-based GF2 calculation due to numerical errors associated with the grid discretization.

The results show the systematic improvability with increasing $\nmom^G$.
The data shows that truncation in $\nmom^G$ at even modest levels leads to a sub-millihartree accuracy in total energy at convergence. For a given $\nmom^G$, the error can be seen to saturate with respect to increasing $\nphys$, with such saturation occurring at larger $\nphys$ for higher $\nmom$. Since these errors are per-electron, this demonstrates that the truncation is size consistent, and that the overall number of auxiliaries required for a given level of accuracy is indeed linear with the size of the system, even in the fully self-consistent approach. Since all physical orbitals remain as canonical Hartree--Fock orbitals, this scaling is not relying on locality arguments, despite the one-dimensional nature of these test systems.

This effective and efficient two-step auxiliary compression algorithm is now used exclusively for the results in the remainder of this work, and we use the notation $AGF2(\nmom^G,\nmom^{\Sigma})$ to denote an auxiliary-GF2 calculation performed with the two-step auxiliary truncation, first to order $\nmom^{\Sigma}$ in the hole/particle self-energy moments, and then to order $\nmom^G$ in the hole/particle Green's function moments, with $GF2$ referring to a grid-based implementation described in Sec.~\ref{sec:GF2}. 

\subsection{Energy Functionals}\label{sec:energy_functionals}
The total energy can be computed directly from the physical and auxiliary systems via the Migdal-Galitskii functional at zero temperature\cite{MigdalGalitskii,PhysRevB.63.075112,PhysRevB.86.081102,Phillips2014}, as 
\begin{equation}
    E = \frac{1}{2 \pi i} \int_{-\infty}^{\infty} \Tr[G(\omega) \Sigma(\omega)]e^{i\omega \eta} d\omega .
\end{equation}
Splitting this into a one-body and two-body part, the one-body contribution to the energy is calculated using the familiar Hartree--Fock functional as
\begin{equation}\label{eq:one_body_energy}
    \eoneb = \frac{1}{2} \Tr [D_\mathrm{phys} (h + F_\mathrm{phys}) ] + E_\mathrm{nuc},
\end{equation}
where $D_\mathrm{phys}$ is the correlated (sub-idempotent) density matrix computed from the projection of the QMOs, as given in Eq.~\ref{eq:Dphys}, $F_\mathrm{phys}$ is defined in Eq.~\ref{eq:Fock} and $E_\mathrm{nuc}$ is the nuclear--nuclear repulsion energy.

The two-body energy is found from the equation of motion form of the Green's function\citep{PhysRevB.63.075112,PhysRevB.86.081102} and is commonly calculated on contours over the Matsubara axis, as
\begin{equation}\label{eq:two_body_energy_se_gf}
    \etwob = \frac{1}{2} \Tr \frac{1}{2\pi} \int_{-\infty}^{\infty} \Sigma (i\omega) G (i\omega) \ d\omega .
\end{equation}
%
%
The expression for the Green's function can be written on the Matsubara axis in terms of the QMOs as
\begin{equation}\label{eq:greens_function}
    [G(i\omega)]_{pq} = \sum_{w}^{\nqmo} \frac{\phi_{pw} \phi_{qw}}{i\omega - \lambda_{w}} .
\end{equation}
where $\lambda,\phi$ are the eigenvalues and eigenvectors of $F_\mathrm{ext}$, respectively. By substituting the expression for $\Sigma(i\omega)$ and $G(i\omega)$ in terms of auxiliaries given by Eq.~\ref{eq:aux_self_energy} and Eq.~\ref{eq:greens_function}, we can analytically integrate Eq.~\ref{eq:two_body_energy_se_gf} to arrive at a frequency-independent, sum-over-states expression for the two-body energy (see Appendix~\ref{app:energy_functional} for derivation) as
\begin{equation}\label{eq:two_body_energy_aux}
    \etwob = 2 \sum_{pq}^{\nphys} \sum_{\alpha}^{\nocc^{\textrm{aux}}} \sum_{b}^{\nvir^{\mathrm{QMO}}}
    \frac{v_{p \alpha} v_{q \alpha} \phi_{pb} \phi_{qb}}{\eps_{\alpha} - \lambda_{b}},
\end{equation}
where $\alpha$ runs over occupied auxiliaries (i.e. auxiliaries with $\eps_{\alpha} < \mu$, representing the poles of hole $\Sigma$) and $b$ runs over virtual QMOs (i.e. poles of the particle $G$). As shown in the Appendix, an equivalent expression is obtained if $\alpha$ runs over virtual auxiliaries and $b$ over occupied QMOs. 

For the MP2 energy, the correlated $G(i\omega)$ is substituted for the Hartree--Fock $G_0(i\omega)$, the zeroth-order Green's function\citep{Holleboom1990},
given by
%
%
%
\begin{equation}\label{eq:hf_greens_function}
    [G_{0}(i\omega)]_{pq} = \frac{n_{p} \delta_{pq}}{i\omega-\eps_{p}} + \frac{\tilde{n}_{q} \delta_{pq}}{i\omega-\eps_{q}},
\end{equation}
where $n_{p}$ is the occupation number of canonical Hartree--Fock (physical) orbital $p$, and $\tilde{n}_p$ is $1-n_{p}$ for unrestricted references or $2-n_{p}$ for restricted references. With this substitution, the integration for the two-body MP2 energy becomes
%
\begin{equation}\label{eq:mp2_energy_aux}
    \emptwo = \sum_{i}^{\nocc^{\mathrm{HF}}} \sum_{\alpha}^{\nvir^{\mathrm{aux}}} \frac{v_{i \alpha}^{2}}{E_{i}-\eps_{\alpha}} ,
\end{equation}
where $i$ denotes occupied Hartree--Fock MOs (i.e. poles of $G_0$), and $\alpha$ runs over virtual auxiliaries, which upon substitution of the relevant expressions from Sec.~\ref{sec:aux_params} reduces to the standard MP2 energy expression.

\subsection{Overview of self-consistent algorithm}\label{sec:self_consistent_algo}
\begin{figure}[H]
    \centering
\includegraphics[trim={0 0 0 0},width=1.0\columnwidth]{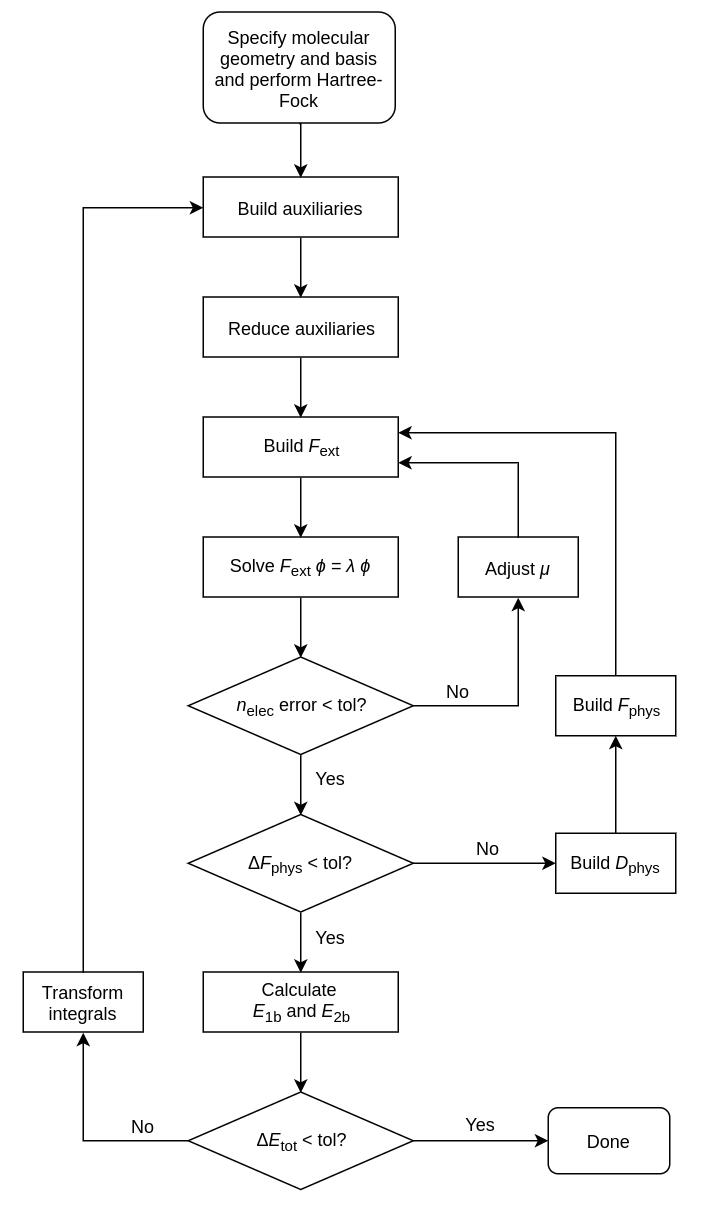}
    \caption{Self-consistent flow chart for the iterative AGF2 procedure.}
    \label{fig:agf2_cycle}
\end{figure}

The overall iterative procedure outlined here is summarized in Fig.~\ref{fig:agf2_cycle}, whereby the original molecular orbitals of the system are self-consistently renormalized by the inclusion of a fictitious auxiliary space. The auxiliary space is built as defined in Sec.~\ref{sec:aux_params} with a cubically growing number, while they are then subsequently compressed to a linear number of auxiliaries with respect to system size in the methods described in Sec.~\ref{sec:compression_schemes}, resulting in a rigorously $\mathcal{O}[N^5]$ algorithm. Energy expectation values with the auxiliary representation are used as derived in Sec.~\ref{sec:energy_functionals}. \toadd{The method typically converges on the scale of tens of iterations, with this dependent on the computational setup used, i.e. degree of damping.} We now turn to the accuracy of the resulting approach in the next section, 
benchmarking the self-consistent auxiliary compression scheme against alternate perturbative approaches, and analyzing its convergence with respect to the number of moments\cite{Phillips2014}. The AGF2, GF2 and OO-MP2\cite{psi4numpy} results were implemented in our own code, making extensive use of existing functionality within the PySCF package\cite{pyscf}.

\section{Results \& Discussion}\label{sec:results_and_discussion}
\subsection{Dissociation Curves}
\begin{figure}
    \centering
    \includegraphics[trim={0 0 0 0},width=1.\columnwidth]{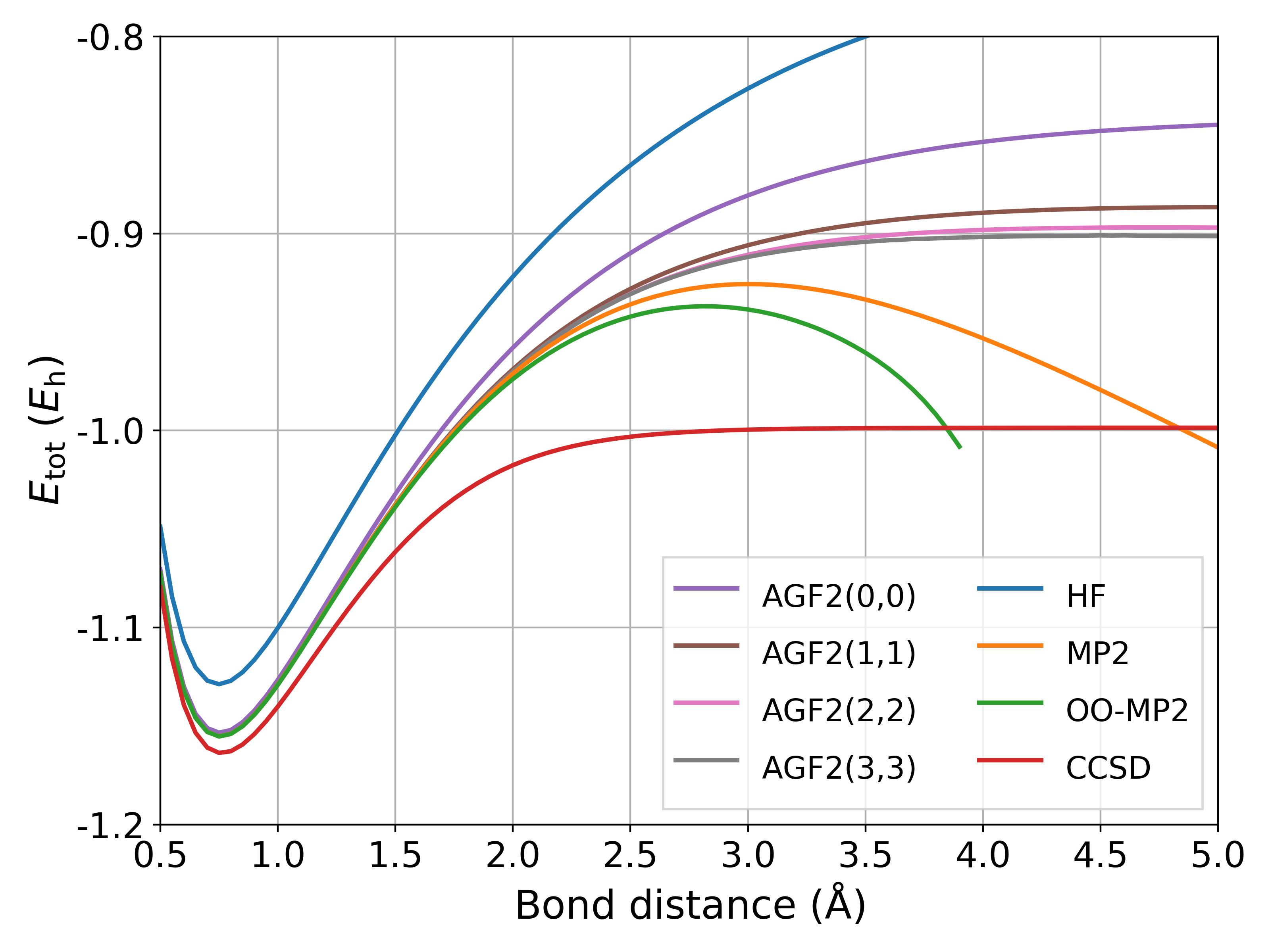}
    \caption{Dissociation of a H$_2$ molecule with a cc-pVDZ basis for a number of methods with a restricted reference.}
    \label{fig:h2_dissociation}
\end{figure}

The dissociation of a single bond is a stern test to demonstrate the transition from a relatively weakly to strongly correlated system, and the ability to correctly break an electron pair\citep{doi:10.1021/bk-2007-0958.ch005}. The paradigmatic example of this is the breaking of the H$_2$ molecule, where in the absence of symmetry-breaking simple perturbative methods such as MP2 describe the equilibrium distance generally well but fail catastrophically upon stretching, while mean-field methods including density functional theory will also make significant errors\citep{doi:10.1021/cr200107z}. This will show the accuracy of the AGF2 for both its limiting dynamical behaviour, as well as the convergence of the auxiliary compression scheme.

Figure~\ref{fig:h2_dissociation} shows dissociation curves for H$_2$ with a cc-pVDZ basis. Shown are Hartree--Fock, MP2, CCSD, and OO-MP2 and a number of moment truncations of AGF2, all with restricted HF references. AGF2 are labelled with $(\nmom^G,\nmom^{\Sigma})$, denoting the effective Green's function and initial self-energy moment truncation based on the scheme in Sec.~\ref{sec:compression_overview}. Coupled-cluster is exact for this two-electron system, and clearly shows the divergent behaviour of MP2 upon stretching. Orbital-optimized MP2, which attempts to perform some refinement of the underlying mean-field spectrum actually diverges faster than the original MP2. In both of these cases, the reason is due to the narrowing of the HOMO-LUMO gap, leading to divergences from the near-singular energy denominator, as has been noted in previous work\cite{Stuck2013}.

However, this closing gap is simply due to the deficiency in the mean-field method in describing the true spectrum of the system in the presence of correlated physics. The true gap is the difference between the ionization potential and the electron affinity, which in the dissociated limit can be approximated by the ionization potential of the Hydrogen atom, equal to 0.5 $E_{\textrm{h}}$ -- a significant gap\cite{doi:10.1063/1.4921259}. The AGF2 provides a set of correlation diagrams to renormalize this spectrum, and in doing so, the auxiliary space hybridizes with the physical RHF orbitals to induce a gap, and a finite and stable total energy even in the dissociation limit. In the case of H$_2$ at 18.0 $\mathrm{\AA}$ the AGF2 gap is found to be a more accurate 0.46$E_\mathrm{h}$, as opposed to the RHF gap of 0.03$E_\mathrm{h}$. Despite this improvement, the energy in this limit is found to substantially overestimate the true dissociation energy, however the shape of the dissociation and flat convergence profile (removing the spurious mean-field tail\cite{doi:10.1021/cr200107z}) is qualitatively accurate opposed to the parent MP2 method\cite{Phillips2014}. Furthermore, the auxiliary truncations with respect to moment order are found to rapidly and systematically converge the total energy, even at low orders. However, we do not include direct comparison to a more traditional grid-based GF2 implementation at this stage, as while it is indistinguishable from AGF2(3,3) for this system for the majority of the binding coordinate, beyond a certain bond length the grid-based GF2 implementation discontinuously jumps to a different solution. Future work will include a more extensive comparison between the $\nmom^G$ truncation of dynamical effects and the traditional fixed grid effective truncation of dynamical resolution in self-consistent Green's function methods, and their relative abilities for systematic and robust convergence, particularly in cases of changing character of the electronic structure.

%
\begin{figure}
    \centering
    \includegraphics[trim={0 0 0 0},width=1.\columnwidth]{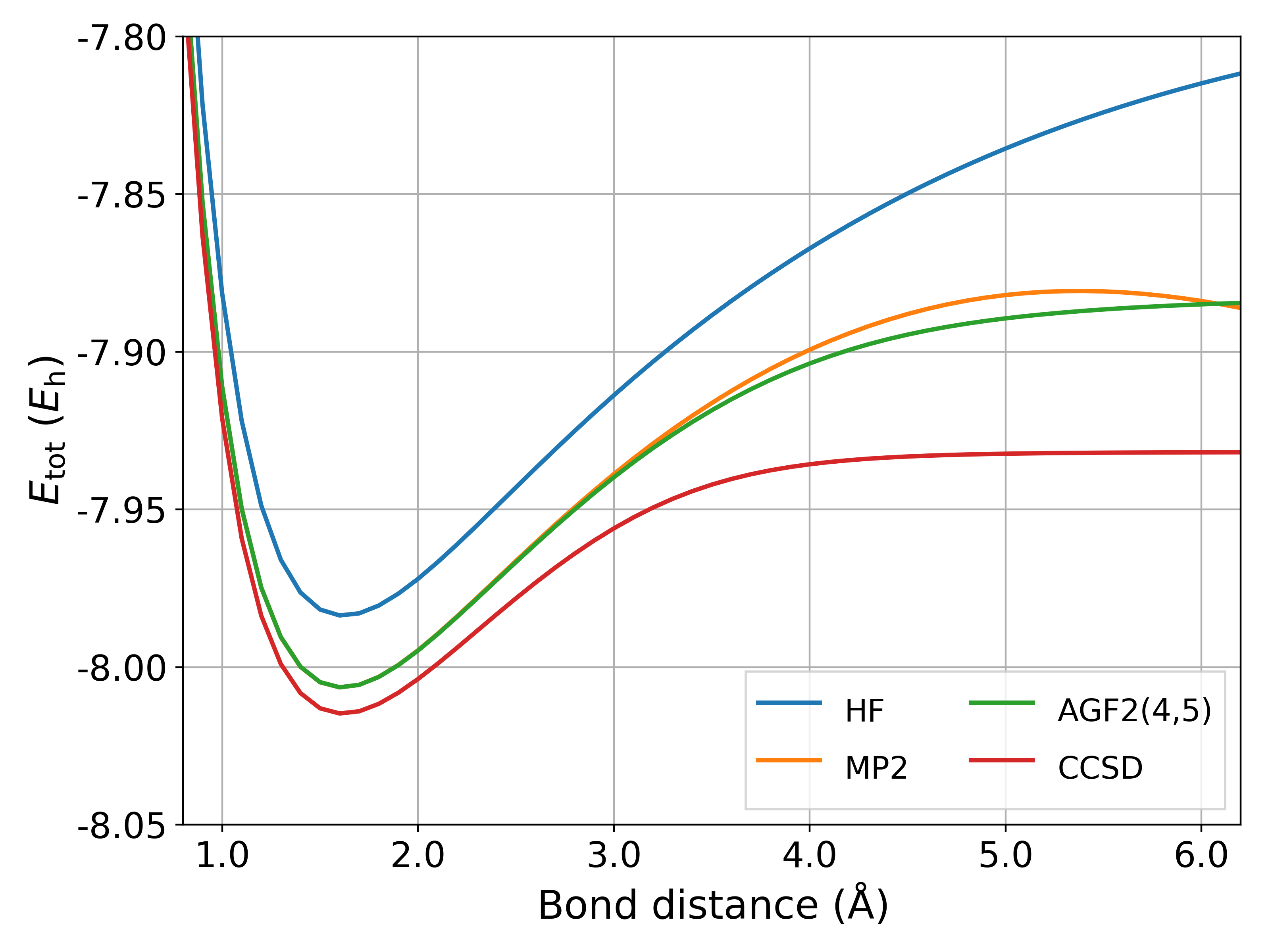}
    \caption{Dissociation of a LiH molecule with a cc-pVDZ basis for a number of methods with a restricted reference.}
    \label{fig:lih_dissociation}
\end{figure}

\begin{figure}
    \centering
    \includegraphics[trim={0 0 0 0},width=1.\columnwidth]{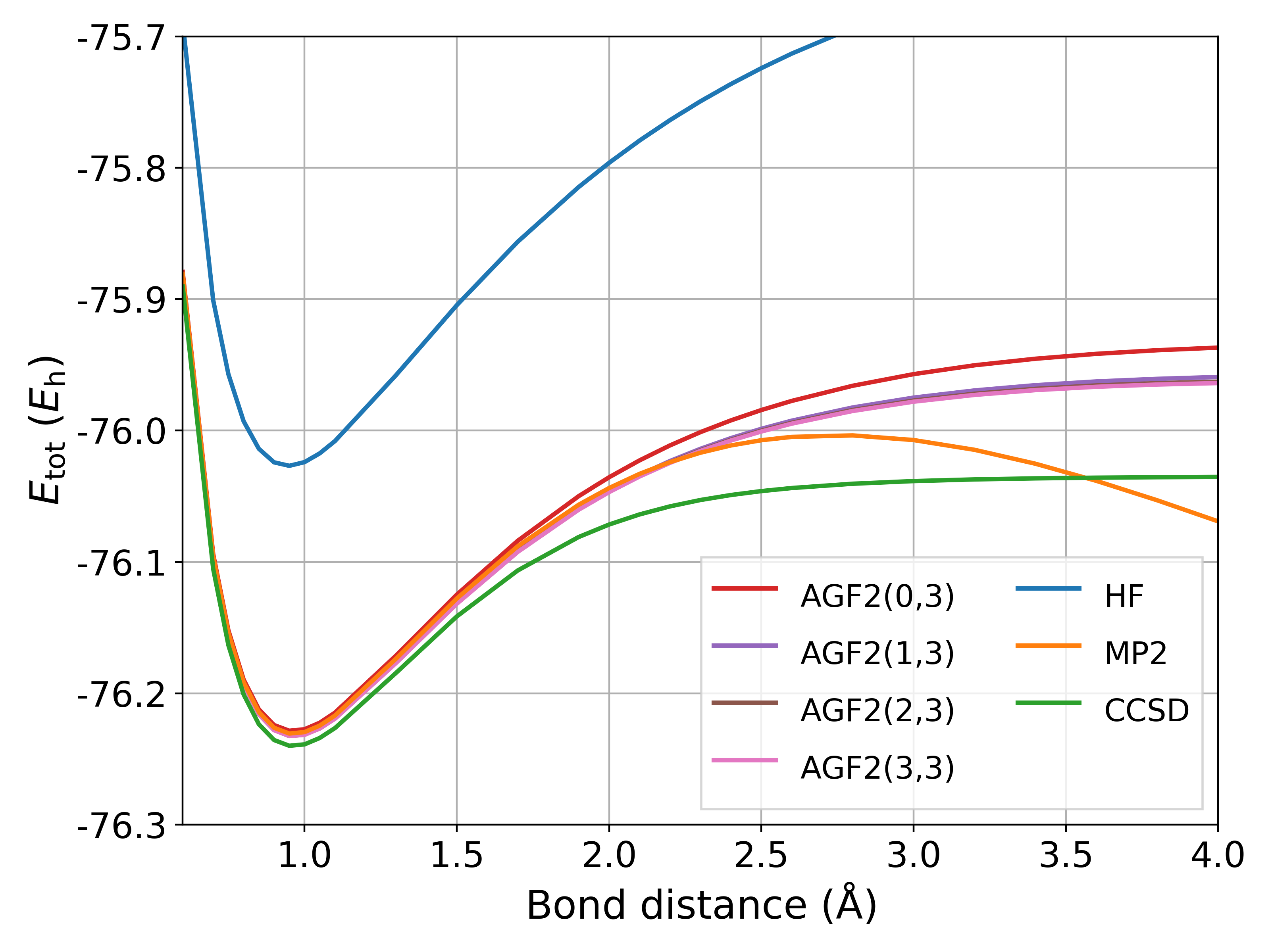}
    \caption{Dissociation of a single H atom from a H$_2$O molecule with a cc-pVDZ basis with increasing $\nmom^G$ truncation with a restricted reference.}
    \label{fig:h2o_dissociation}
\end{figure}

We also show dissociation profiles for LiH (Fig.~\ref{fig:lih_dissociation}) and of a single Hydrogen atom extraction from a water molecule (Fig.~\ref{fig:h2o_dissociation}) in cc-pVDZ basis sets. \toadd{In these cases, CCSD is not exact and therefore CCSD(T) is shown instead.} Once again, the results are essentially identical to MP2 close to equilibrium, while avoiding the divergences which plague MP2 as stronger correlation effects are introduced. It is also found that a $\nmom^G$ order as low as 1 is sufficient to achieve highly accurate values for the energies compared to the converged moment limit.
This indicates that the modification of the variance and skew of the particle and hole spectral moments (the second and third moments of these distributions, as faithfully maintained by $\nmom^G=1$) is sufficient for accurate energetics in these systems both at equilibrium and strongly correlated limits. This results in the relation $\naux \leq 3\nphys$, meaning that the highest scaling step (the integral transform of Eq.~\ref{eq:integral_transform}) has a prefactor of only approximately 27 compared to a traditional MP2 integral transform, neglecting density-fitting or other efficiency-related adaptations.

It should be noted that for the stretched geometries of these data, we sometimes observed convergence to alternate self-consistent fixed points,
and it is often difficult to constrain the method to robustly find the same solution at different bond lengths. The existence of multiple valid stationary points is a feature well known in iterative Green's function theories\citep{PhysRevLett.114.156402}, with this behaviour also common within the grid-based GF2 method and without a variational principle to rely on. We therefore do not consider this behaviour to be a result of the auxiliary state formulation of the method. 
The data chosen here do not display this behaviour, and in situations where this can be a problem one may potentially make small changes to the computational set up to combat it, such as changing the truncation order, or performing damping of the auxiliaries through the iterations.

\subsection{G1 Set}
\begin{figure*}
    \centering
\includegraphics[trim={0 0 0 0 }, width=2.\columnwidth]{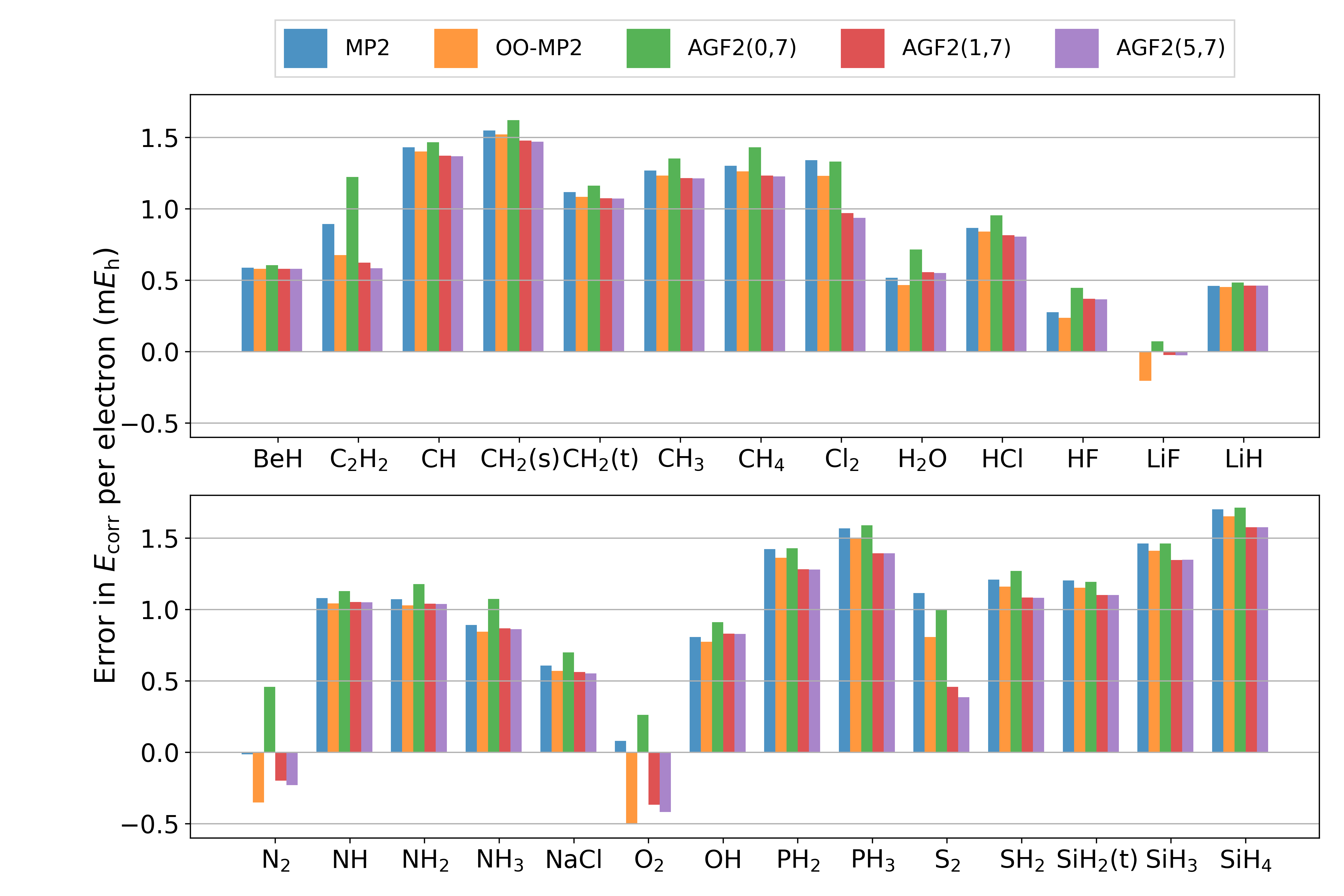}
    \caption{Comparison of errors in the all-electron correlation energy per electron for MP2, OO-MP2 and three moments of $\nmom^G$ truncation (0, 1 and 5), each with an initial truncation to $\nmom^\Sigma = 7$. Error is taken with respect to the CCSD(T) energy, with all calculations performed using an unrestricted reference. Data consists of a subset of the G1 test set of molecules, with a cc-pVDZ basis, using an unrestricted reference, with molecular geometries optimized at the MP2/aug-cc-pVDZ level. 
    \toremove{Singlet and triplet states are distinguished by (s) and (t), respectively.}
    \toadd{Triplet states are marked with (t), where all other molecules are singlets or doublets.}}
    \label{fig:g1_error}
\end{figure*}

\begin{table}
    \centering
    \begin{tabular}{l@{\hskip 1cm}c@{\hskip 1cm}c@{\hskip 1cm}}
        \hline
         & \multicolumn{2}{c@{\hskip 1cm}}{Error per electron (m$E_\mathrm{h}$)} \\
        Method & Mean & Maximum\\
        \hline
        MP2 & 1.175 & 1.850 \\
        OO-MP2 & 1.084 & 1.800 \\
        AGF2(1,7) & 1.062 & 1.725 \\
        AGF2(5,7) & 1.051 & 1.725 \\
        \hline
    \end{tabular}
    \caption{Mean and maximum absolute errors in mE$_\mathrm{h}$ in the correlation energy per electron for MP2, OO-MP2, AGF2(1,7) and AGF2(5,7) compared to CCSD(T). Data corresponds to that of Fig.~\ref{fig:g1_error}. The maximum errors correspond to SiH$_4$ in the case of all three methods shown.}
    \label{tab:g1_error_values}
\end{table}

In order to consider a wider selection of molecular systems of interest at equilibrium geometry, Fig.~\ref{fig:g1_error} shows results from calculations on the total energies for the G1 set of molecules\citep{doi:10.1063/1.456415,doi:10.1063/1.458892,doi:10.1063/1.4959245}. These were performed with all electrons correlated in a cc-pVDZ basis, and compared to the CCSD(T) energy which was taken as the benchmark single-reference method. This set includes a number of systems which are open-shell or non-zero spin states, which required an adaptation for unrestricted AGF2 formulation where the auxiliary states couple differently to the two spin sectors of the physical orbitals. In the plot, we compare the correlation energy error per electron for each system for MP2, OO-MP2 and AGF2 with a truncation of $\nmom^G=0, 1$ and $5$, to examine the larger trends in accuracy with increased truncation. Furthermore, there were a small number of systems for which the AGF2 iterations for these truncations clearly and obviously converged to an incorrect and unphysical solution, often being in error energetically on the order of several tenths of hartrees, as described in the previous section. These unphysical solutions were easily identified and omitted, so as not to detract from the overall trends in the accuracy of the method in the absence of convergence errors.

While these molecules at their equilibrium geometries are not to be strongly correlated and so large-scale improvements are not expected, there are some trends which are evident. Firstly, the AGF2(5,7) results, which are effectively converged to the large moment limit, show a relatively small but appreciable improvement over the MP2 and OO-MP2 results across the test set. These aggregated errors are shown in Tab.~\ref{tab:g1_error_values} giving the mean and maximum absolute deviation, demonstrating a consistent $\sim 0.12$ m$E_{\textrm h}$ per electron improvement in both the average and worst-case results. This improvement is largest for the heavier systems including P, S or Si compounds hinting that the presence of small amounts of stronger correlation effects may be responsible for the largest improvements in the electronic description compared to MP2. This is most clearly seen in the S$_2$ molecule, where the error per electron to CCSD(T) drops from 1.706~m$E_\textrm{h}$ for MP2 to 0.977~m$E_\textrm{h}$ for AGF2(5,7).
The improvement between MP2 and OO-MP2 is far smaller, indicating that the stationarity of the MP2 functional with respect to orbital rotations is not a sufficient condition to correctly modify the underlying propagator to substantially improve these generally single-reference systems.

Another noteworthy feature of these results is the consistency in the results between $\nmom^G=1$ and $\nmom^G=5$, which only have an average deviation of 0.011 m$E_\textrm{h}$ between them. This demonstrates the rapid convergence of the errors with $\nmom^G$ truncation, with again $\nmom^G=1$ (representing a faithful representation of up to the third spectral moment of the occupied and virtual states of ${\hat H}_0$) being responsible for the vast majority of the improvement compared to MP2. However this improvement is not found for $\nmom^G=0$, which generally performs worse than the MP2 result. This truncation is representative of the changes in the underlying physical space one-body density matrix and Fock matrix in response to the correlations, and these changes appear unable in isolation to improve upon MP2, indicating that the more significant changes to the underlying spectrum provided by $\nmom^G=1$ are required.
\toremove{The exception of this is seen in N$_2$ and O$_2$, which represent the only systems studied where some calculated correlation energies are below those of CCSD. These systems also correspond to those where MP2 already performs exceptionally well. Despite in these cases $\nmom^G=5$ being further away from CCSD than $\nmom^G=1$, they still follow the same general trends: OO-MP2 provides a larger correlation energy than MP2, with AGF2(0,7) providing substantially less correlation energy than OO-MP2.} There is a much smaller difference between the $\nmom^G=1$ and $\nmom^G=5$ AGF2 solutions than that between $\nmom^G=0$ and $\nmom^G=1$, showing that the significant portion of the convergence of the energetics with respect to $\nmom^G$ comes from increasing $\nmom^G = 0$ to $1$, and not from including the more extensive dynamical character of the high-order moment expansion.


\section{Conclusions}\label{sec:conclusions}
We have presented a renormalized second-order perturbation theory which allows for the modification of the spectrum of ${\hat H}_0$ in response to the MP2 correlations. As opposed to traditional GF2, the method is cast entirely in terms of static quantities and avoids the specification and sensitivity of grids and numerical Fourier transforms. The self-consistent self-energy is implicit, and instead represented directly in terms of non-interacting auxiliary states which couple to the physical orbitals of the system. We have demonstrated a rigorous $\mathcal{O}[N^5]$ scaling of the resulting method, with the presentation of a novel truncation scheme to compact the auxiliary space representation based on rigorously conserving the occupied and virtual spectral moments of the resulting correlated spectrum. The results demonstrate that the truncation to just the third spectral moment ($\nmom^G=1$) is already a significant improvement on the simple MP2 (or orbital-optimized MP2) results, both in weakly correlated, but more substantially for strongly correlated systems where the renormalization of the spectrum can prevent divergences which plague MP2. This truncation results in the requirement of an auxiliary space less than three times the size of the physical space. Furthermore, the method also holds promise to exploit locality in the auxiliary space in order to further reduce the formal scaling of the method, which will be investigated in the future.

\section{Acknowledgements}

The authors thank Dominika Zgid for helpful discussions. G.H.B. also gratefully acknowledges support from the Royal Society via a University Research Fellowship, as well as funding from the European Union's Horizon 2020 research and innovation programme under grant agreement No. 759063.

\section{Appendix: Sum-over-states two-body energy functional} \label{app:energy_functional}

We begin with the Galitskii--Migdal formula for the two-body energy, expressed as a sum over physical degrees of freedom upon the Matsubara axis. Assuming a restricted reference:
\begin{align*}
    E_\mathrm{2b}
    &= \frac{1}{2} \frac{1}{2\pi} \sum_{\sigma}^{n_\mathrm{spin}} \sum_{pq}^{n_\mathrm{phys}} \int _{-\infty}^{\infty} [\Sigma_\mathrm{MP2}^{\sigma} (i\omega)]_{pq} [G^{\sigma} (i\omega)]_{pq} d\omega
    \\
    &= \frac{1}{2\pi} \sum_{pq}^{n_\mathrm{phys}} \int _{-\infty}^{\infty} [\Sigma_\mathrm{MP2} (i\omega)]_{pq} [G (i\omega)]_{pq} d\omega
    \\
    &= \frac{1}{2\pi} \sum_{pq}^{n_\mathrm{phys}} \sum_{\alpha}^{n_\mathrm{aux}} \sum_{w}^{n_\mathrm{QMO}} v_{p\alpha} v_{q\alpha} \phi_{pw} \phi_{qw} \\
    &\quad \times \int _{-\infty}^{\infty} \left( \frac{1}{i\omega - \epsilon_{\alpha}} \right) \left( \frac{1}{i\omega - \lambda_{w}} \right) d\omega
    ,
\end{align*}
where $\lambda,\phi$ are the poles of the Green's function, obtained via diagonalisation of the extended Fock matrix.
The last integral can be rewritten as
\begin{equation*}
    - \int _{-\infty}^{\infty}
    \left( \frac{1}{\omega + i \epsilon_{\alpha}} \right)
    \left( \frac{1}{\omega + i \lambda_{w}} \right) d\omega
\end{equation*}
and evaluated via contour integration.
If both $\epsilon_\alpha > 0$ and $\lambda_w > 0$, the poles of the integrand $-i\epsilon_\alpha$ and
$-i \lambda_w$ will lie in the lower half-plane and the contour can be closed with a semi-circle of infinite radius in the upper half-plane.
Since the closed contour does not encircle any poles, the integral is zero.
The semi-circle at infinite distance to the origin does not contribute to the integral, since the integrand disappears faster than $\frac{1}{\omega}$ and so the line integral along the real axis in itself must be zero.
The same holds for $\epsilon_\alpha < 0$ and $\lambda_w < 0$, since the contour can be closed in the lower half-plane.
The line integral is thus only non-zero if $\epsilon_\alpha > 0$ and $\lambda_w < 0$ or vice versa.
In the former case, closing the contour in the upper half-plane only encircles the pole at $\omega = -i \lambda_w$, which has a residue of $i / (\epsilon_\alpha - \lambda_w)$.
The contour integral and hence line integral thus become $2 \pi / (\lambda_w - \epsilon_\alpha)$.
Similarly, the case $\epsilon_\alpha > 0$ and $\lambda_w < 0$ results in an integral value of $2 \pi / (\epsilon_\alpha - \lambda_w)$.
We can thus write
\begin{equation*}
\begin{split}
    \int _{-\infty}^{\infty} \left( \frac{1}{i\omega - \epsilon_{\alpha}} \right)
    \left( \frac{1}{i\omega - \lambda_{w}} \right) d\omega \\ =
    \begin{cases}
    2 \pi / (\lambda_w - \epsilon_\alpha) & \mathrm{if} \: \epsilon_\alpha > 0, \lambda_w < 0  \\
    2 \pi / (\epsilon_\alpha - \lambda_w) & \mathrm{if} \: \epsilon_\alpha < 0, \lambda_w > 0  \\
    0 & \mathrm{else}
    \end{cases}
\end{split}
\end{equation*}
and restrict the two-body energy expression to a sum-over-states expression in terms of occupied $\epsilon_\alpha$ and virtual
$\lambda_w $ and vice versa:
\begin{align*}
    E_\mathrm{2b}
    = \sum_{pq}^{n_\mathrm{phys}}
    & \left( \sum_{\alpha}^{n^\mathrm{aux}_\mathrm{occ}} \sum_{b}^{n^\mathrm{QMO}_{\mathrm{vir}}}
    \frac{v_{p\alpha} v_{q\alpha} \phi_{pb} \phi_{qb}}{\epsilon_\alpha - \lambda_b}
    \right.
    \\
    + & \left.\sum_{\alpha}^{n^\mathrm{aux}_\mathrm{vir}} \sum_{i}^{n^\mathrm{QMO}_{\mathrm{occ}}}
    \frac{v_{p\alpha} v_{q\alpha} \phi_{pi} \phi_{qi}}{\lambda_i - \epsilon_\alpha}
    \right)
    .
\end{align*}
We note without proof that the first and second terms are equal and the equation thus simplifies to
\begin{equation*}
    E_\mathrm{2b}
    =  2 \sum_{pq}^{n_\mathrm{phys}}
    \sum_{\alpha}^{n^\mathrm{aux}_\mathrm{occ}} \sum_{b}^{n^\mathrm{QMO}_{\mathrm{vir}}}
    \frac{v_{p\alpha} v_{q\alpha} \phi_{pb} \phi_{qb}}{\epsilon_\alpha - \lambda_b}
    .
\end{equation*}

If one uses $G_0$ in place of $G$ and follows the equivalent derivation, the non-zero contributions become
\begin{align*}
    E_\mathrm{2b}^{(0)} 
    &= 2 \sum_{p}^{n_\mathrm{occ}^\mathrm{phys}} \sum_{\alpha}^{n_\mathrm{vir}^\mathrm{aux}} \frac{v_{p\alpha}^{2}}{\epsilon_{p} - \epsilon_{\alpha}} ,
\end{align*}
which, upon substituting the expressions for the auxiliaries, is twice the MP2 energy. This is discussed by Holleboom and Snijders\cite{Holleboom1990} and we can therefore write the MP2 energy in terms of the poles as
\begin{align*}
    E_\mathrm{MP2}
    &= \sum_{i}^{n_\mathrm{occ}^\mathrm{HF}} \sum_{\alpha}^{n_\mathrm{vir}^\mathrm{aux}} \frac{v_{i\alpha}^{2}}{E_i - \epsilon_{\alpha}},
\end{align*}
where $E_i$ now denotes the occupied Hartree--Fock eigenvalues.

\section{References}

\bibliographystyle{achemso}
\providecommand{\latin}[1]{#1}
\makeatletter
\providecommand{\doi}
  {\begingroup\let\do\@makeother\dospecials
  \catcode`\{=1 \catcode`\}=2 \doi@aux}
\providecommand{\doi@aux}[1]{\endgroup\texttt{#1}}
\makeatother
\providecommand*\mcitethebibliography{\thebibliography}
\csname @ifundefined\endcsname{endmcitethebibliography}
  {\let\endmcitethebibliography\endthebibliography}{}

\end{document}